\documentclass[final,5p,times,twocolumn]{elsarticle}

\usepackage{graphicx}

\usepackage{amssymb}
\usepackage{amsmath}
\usepackage{booktabs, supertabular, ctable}
\providecommand{\abs}[1]{\lvert#1\rvert}

\newcommand{\cm}{cm$^{-1}$}

\journal{Chemical Physics Letters}

\begin{document}

\begin{frontmatter}



\title{Broadband velocity modulation spectroscopy of HfF$^+$: towards a measurement of the electron electric dipole moment}


\author[jila]{Kevin C. Cossel\corref{cor1}}
\cortext[cor1]{Corresponding Author}
\ead{kevin.cossel@colorado.edu}
\author[jila]{Daniel N. Gresh}
\author[jila]{Laura C. Sinclair\fnref{fn1}}
\fntext[fn1]{Current address: National Institute of Standards and Technology, 325 Broadway, Boulder, CO}
\author[jila]{Tyler Coffey}
\author[petersburg1,petersburg2]{Leonid V. Skripnikov}
\author[petersburg1,petersburg2]{Alexander N. Petrov}
\author[petersburg1]{Nikolai S. Mosyagin}
\author[petersburg1]{Anatoly V. Titov}
\author[mit]{Robert W. Field}
\author[jila]{Edmund R. Meyer\fnref{fn2}}
\fntext[fn2]{Current address: Department of Physics, Kansas State University, Manhattan, KS 66506}
\author[jila]{Eric A. Cornell}
\author[jila]{Jun Ye}
\address[jila]{JILA, National Institute of Standards and Technology and University of Colorado, and\\
Department of Physics University of Colorado, 440 UCB, Boulder, CO 80309, USA}
\address[petersburg1]{Petersburg Nuclear Physics Institute, Gatchina, Leningrad district 188300, Russia}
\address[petersburg2]{Department of Physics, St. Petersburg State University, 198904, Russia}
\address[mit]{Department of Chemistry, MIT, Cambridge, MA 02139, USA}

\begin{abstract}

Precision spectroscopy of trapped HfF$^+$ will be used in a search for the permanent electric dipole moment of the electron (eEDM). While this dipole moment has yet to be observed, various extensions to the standard model of particle physics (such as supersymmetry) predict values that are close to the current limit. We present extensive survey spectroscopy of 19 bands covering nearly 5000~cm$^{-1}$ using both frequency-comb and single-frequency laser velocity-modulation spectroscopy. We obtain high-precision rovibrational constants for eight electronic states including those that will be necessary for state preparation and readout in an actual eEDM experiment.


\end{abstract}


\end{frontmatter}


\section{Introduction}
\label{introduction}

The existence of a permanent electric dipole moment of an electron (eEDM) would have profound implications for fundamental physics since it violates parity and time reversal symmetries \cite{Hinds1997, Sandars2001, Fortson2003}. The current experimental limit of about $1\times10^{-27}$~$e\times cm$ \cite{Regan2002, Hudson2011} is many orders of magnitude larger than the predicted eEDM from the standard model \cite{Commins1999}, but extensions to the standard model (such as supersymmetry) predict a dipole moment on the order of $10^{-29}$ to $10^{-26}$~$e\times cm$ \cite{Commins1999}. This means that measurement of the eEDM provides a rigorous test of extensions to the standard model without having to disentangle new results from standard model predictions. HfF$^+$ and the related ThF$^+$ have been suggested as candidate species in such a measurement using trapped molecular ions \cite{Meyer2006, Meyer2008, Petrov2007, Leanhardt2011}. The low-lying, metastable $^3\Delta_1$ states in HfF$^+$ and ThF$^+$ have high sensitivity to the eEDM due to the large effective electric field felt by one of the unpaired electrons when the molecule is polarized in low laboratory electric fields. As initially shown by DeMille \textit{et al} for PbO, the level structure of $\Omega = 1$ states provides valuable checks to reduce systematic errors \cite{DeMille2001}. The use of molecular ions makes trapping straightforward, thus enabling measurements with long coherence times. 

\begin{figure}[t]
\includegraphics[scale=0.95]{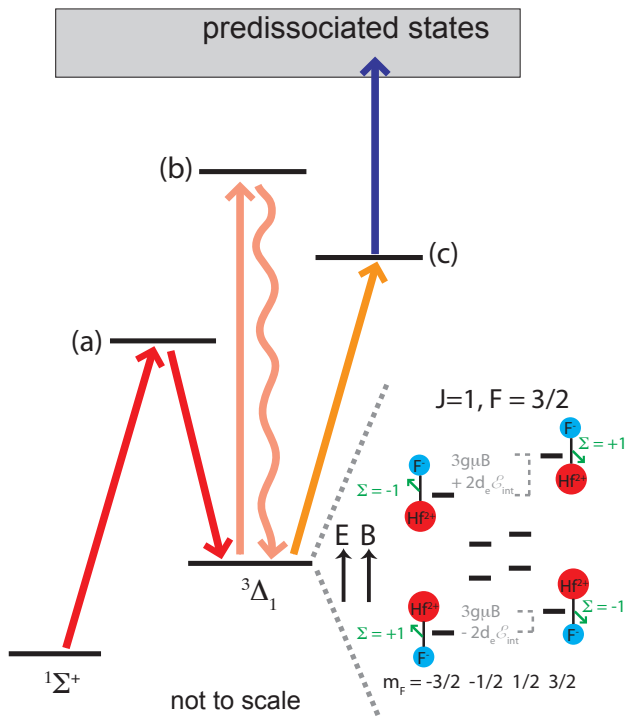}
\caption{\textbf{Energy levels of HfF$^+$ relevant to the eEDM experiment.} The eEDM experiment will utilize the J = 1 level of the $^3\Delta_1$ metastable state. Mixed-spin excited states that couple well to both the ground $^1\Sigma^+$ and the $^3\Delta_1$ could be used for population transfer (state (a)). Readout from the $^3\Delta_1$ could be accomplished directly via laser-induced fluorescence (b) or resonantly-enhanced multiphoton photodissociation (c). The inset shows the structure of the $J=1$, $F = 3/2$ manifold of the $^3\Delta_1$ level in applied electric and magnetic fields. The eEDM experiment will measure the difference in energy between the upper (and lower) pairs of $\abs{m_F}=3/2$ states. The applied magnetic field splits these states by $3g\mu B$, where $B$ is the magnitude of the magnetic field, $\mu$ is the Bohr magneton, and $g$ is the magnetic g-factor. In addition, the electron experiences an effective internal electric field ($\mathcal{E}_{int}$), which results in an additional splitting with magnitude $2 d_e \mathcal{E}_{int}$, where $d_e$ is the eEDM (aligned with the electron spin).}
\label{energylevels}
\end{figure}

The general approach for the eEDM experiment with HfF$^+$ or ThF$^+$ will be to perform a Ramsey-type spectroscopy measurement of the energy separation of two $\abs{m_F}=3/2$ magnetic sublevels of the $^3\Delta_1$ $J=1$ level (see Figure~\ref{energylevels}). In applied magnetic and electric fields these sublevels are split both by the Zeeman effect and by the interaction of the dipole moment of the electron (assumed to be parallel to the electron spin) with the internal electric field of the ion. This magnitude of this shift is given by $2 d_e \mathcal{E}_{int}$ (using the convention of \cite{Petrov2007} instead of that in \cite{Leanhardt2011}), where $d_e$ is the eEDM to be measured and $\mathcal{E}_{int}$ is the effective internal electric field experienced by the electron (about 24~GeV/cm \cite{Petrov2007}). HfF$^+$ will initially be formed and trapped in the X$^1\Sigma^+$ state; then optical (and possibly microwave) transitions will be used to create a coherent superpostion in the two desired $m_F$ sublevels of the $^3\Delta_1$ state. After a free-evolution period a readout pulse will be applied to meaure the accumulated phase difference between the two sublevels, which is proportional to the energy difference between the sublevels. A detailed analysis of the proposed experiment can be found in \cite{Leanhardt2011}. For this scheme, optical transitions from the metastable $^3\Delta_1$ state will most likely be required for state-selective readout via laser induced fluorescence or resonant multiphoton photodissociation (states (b) or (c) in Figure~\ref{energylevels} respectively); in addition, transitions coupling the X$^1\Sigma^+$ state with the $^3\Delta_1$ state via some excited state (state (a) in Figure~\ref{energylevels}) will be necessary for state preparation in the $^3\Delta_1$ state. 

Prior to this work, very little spectroscopic information was available for either HfF$^+$ or ThF$^+$. The energies of the low-lying $^3\Delta_1$, $^3\Delta_2$, and $^3\Delta_3$ states were recently measured using pulsed-field ionization zero kinetic energy (PFI-ZEKE) spectroscopy \cite{Barker2011, Barker2012}. Additionally, several isoelectronic species including TiO \cite{Kaledin1995, Namiki1998, Kobayashi2002}, TiF$^+$ and TiCl$^+$ \cite{Focsa1998, Focsa1999} have been studied. Importantly though, there are no fully characterized optical transitions in either HfF$^+$ or ThF$^+$ at wavelengths below 1 $\mu$m (above $10^4$~cm$^{-1}$) where laser-induced fluorescence detection is possible. Due to the large number of electrons involved, quantum calculations are challenging: current high-level calculations have errors of perhaps 1000~cm$^{-1}$ \cite{Petrov2007, Petrov2009}.

The combined lack of experimental data and large theoretical uncertainties necessitated precision spectroscopy over a very broad spectral range. To achieve this, we recently developed  frequency-comb velocity-modulation spectroscopy (comb-vms), which provides simultaneously broad bandwidth, high resolution, high sensitivity, and ion specificity \cite{Sinclair2011}. We were able to measure 1000~cm$^{-1}$ of continuous spectra around 12000~cm$^{-1}$ using comb-vms and observed four bands of HfF$^+$. This information, combined with PFI-ZEKE data and theoretical predictions, was then used to guide specifically targeted scans based on single-frequency (cw-laser) velocity-modulation spectroscopy, which resulted in 15 additional bands with origins ranging from 9950 to 14600~cm$^{-1}$. We obtained precise molecular constants for the $^1\Sigma^+$,  $^3\Delta_1$, $^3\Pi_{0^-}$, $^3\Pi_{0^+}$, $^1\Pi_1$, $^3\Pi_1$, $^3\Sigma_{0^+}^-$, and $^3\Phi_2$ states, many of which will be of use for the eEDM experiment. We were also able to measure the electronic contributions to isotope splittings. In addition, we discuss the observed $\Lambda$-doublings, including the first measurement of the size of the small doubling in the $^3\Delta_1$. This value is important for determining the necessary applied electric field to fully polarize HfF$^+$ as well as for estimating other systematic effects in the final eEDM experiment. Finally, the large amount of data enabled refinements in the \textit{ab initio} calculations; we briefly discuss the modifications and compare the new theory with the measurements. This may help to improve the accuracy of other calculations involving high-Z atoms.

Outside of the eEDM experiment, precision spectroscopy of molecular ions is useful in a wide range of fields including fundamental physics, chemistry, and astrophysics. In astrochemistry for example, at least 22 ions (both positive and negative) ranging in complexity from H$_3^+$ to H$_2$COH$^+$ and C$_8^-$ have been identified in interstellar and circumstellar gases \cite{Klemperer2011}. Many of these ions are highly reactive and are believed to be intermediates in a variety of reactions; nonetheless, some rate constants and branching ratios are still not well known \cite{Snow2008}. For example, the pathway for formation of a very simple molecular ion, CH$^+$, remains elusive \cite{Indriolo2010}. Searches for new species, including efforts to identify the origin of the diffuse interstellar bands \cite{Sarre2006, Oka2011}, benefit from laboratory measurements of optical transitions as well as measurements of rotational constants to aid microwave spectroscopy \cite{Snow2006, Halfen2007, Maier2011}. On the physical chemistry side, precision spectroscopy of H$_3^+$ above the barrier to linearity (near 9913~cm$^{-1}$) provides rigorous tests for \textit{ab initio} theory \cite{Gottfried2003, Morong2009, Pavanello2012}. Many other carbocations are interesting for both their presence as intermediates in reactions such as combustion as well as the challenges they present to theory; for example, the spectrum of CH$_5^+$, a highly non-classical carbocation, still remains unassigned due to both its complexity and spectral interference from other contaminant species \cite{White1999, Asvany2005, Huang2006}. A combination of comb-vms with a recently developed ion-beam spectrometer capable of sensitive, sub-Doppler vms of rotationally cold species (\cite{Mills2011}) could enable mass-selective spectroscopy of ions over a broad bandwidth with many potential applications.

\begin{figure}[t]
\includegraphics[scale=0.95]{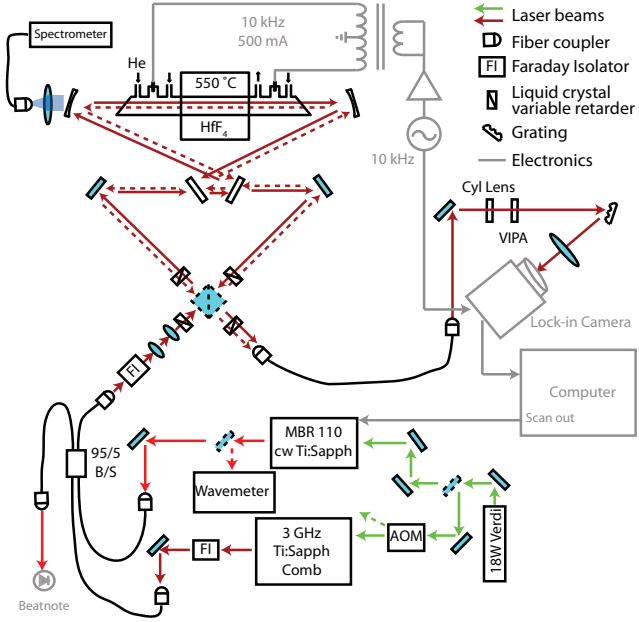}
\caption{\textbf{Setup for comb-vms.} We use a Ti:sapphire comb with a repetition rate of 3 GHz, which is transferred via fiber to the optical cavity. The comb is coupled into one direction of a ring bowtie cavity, which contains an ac-discharge cell with Brewster-angled end windows. The cavity finesse is about 100 and the length is matched to be an integer multiple (25) of the comb cavity. We use liquid crystal variable retarders and a polarizing beamsplitter to alternate the direction of propagation through the cavity. The ac-discharge is driven at 10~kHz with about 500~mA. We flow helium gas through the tube at a pressure of about 6~Torr. HfF$^+$ ions are created by heating about 0.5~g of HfF$_4$ powder to 550 $^\circ$C inside the discharge tube. The cavity transmitted light, which contains the modulated absorption signal, is analyzed using a two-dimensional imaging system and a lock-in camera. A cw Ti:sapphire laser is used to stabilize the comb and serves as a frequency reference in the spectra. More details can be found in \cite{Sinclair2011}. A fiber-coupled grating spectrometer is used to monitor the discharge fluorescence.}
\label{combsetup}
\end{figure}

\begin{figure}[t]
\includegraphics[scale=0.95]{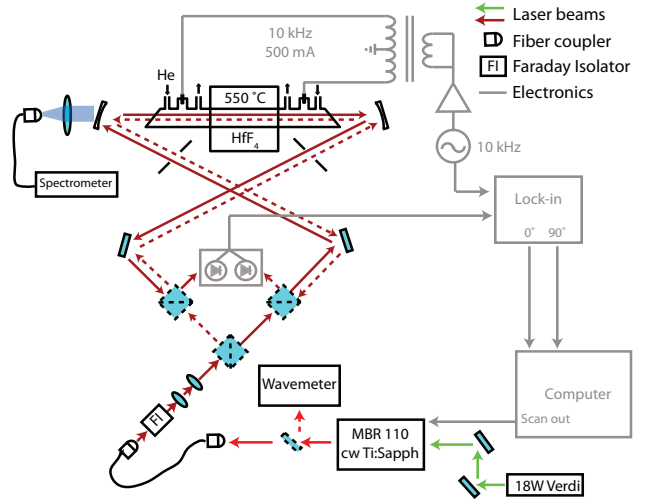}
\caption{\textbf{Setup for cw-vms.} We use a cw Ti:sapphire laser that is tunable from about 9500 - 14800~cm$^{-1}$ (1050 - 675~nm). After fiber coupling between tables, the laser is split with a 50/50 non-polarising beamsplitter. Each beam goes through an additional 50/50 non-polarizing beamsplitter (to separate the return light) and is then sent in opposite directions through the discharge cell. Half of the returning light in each direction is reflected by the beamsplitters and is differenced using an auto-balanced photodetector \cite{Hobbs1997}. This difference signal is then sent to lock-in detectors to record signal in-phase and 90$^\circ$ out-of-phase with the HfF$^+$ absorption. We placed irises in the beam paths to reduce the amount of discharge fluorescence observed on the auto-balanced detector; this helped both to reduce noise and primarily to reduce baseline drifts in the spectra.}
\label{cwsetup}
\end{figure}

\begin{figure}[t]
\includegraphics[scale=0.95]{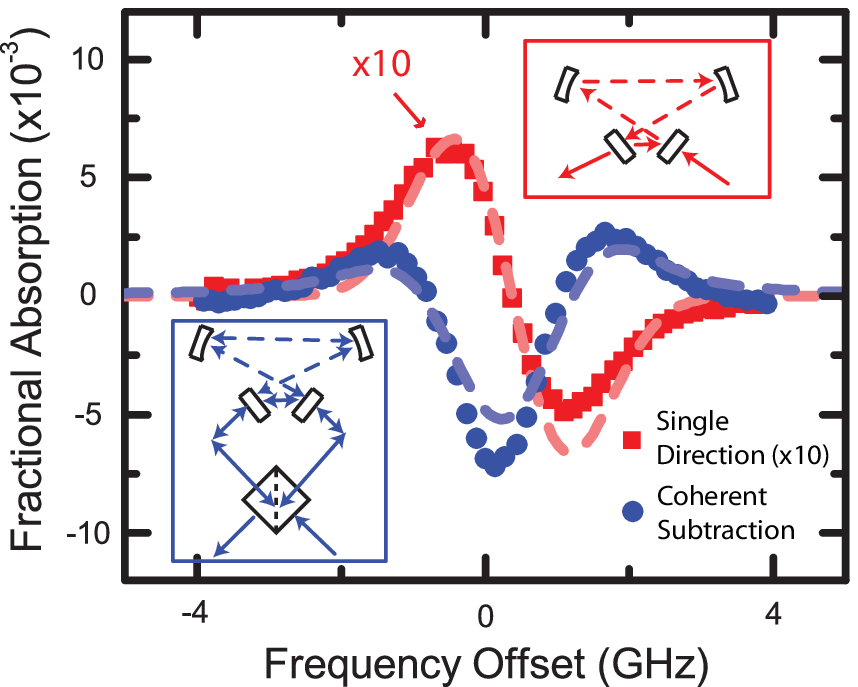}
\caption{\textbf{Comparison of coherent subtraction and single direction measurements.} The signal to noise for coherent subtraction (using a 52/48 beam splitter) surpasses that of single direction measurements by a factor of 10 when the noise is primarily technical light intensity noise.  Both measurements of a single N$_2^+$ line were made with the cw laser.  The dashed lines are a calculation of the expected lineshape (see text).  The modification in the lineshape for the coherent subtraction is due to a stray differential phase shift of approximately $\pi/20$ between the two counter-propagating beams, which results in a sensitivity to both absorption and dispersion.}
\label{coherentsub}
\end{figure}

\section{Experiments}
\label{experiments}
\subsection{Comb-vms}
Comb-vms combines cavity-enhanced direct frequency comb spectroscopy \cite{Thorpe2006, Thorpe2008, Adler2010} with velocity-modulation spectroscopy \cite{Gudeman1983, Saykally1988a, Lindsay2001, Stephenson2005} for discrimination between ions and neutral species. This technique could also be readily extended to the detection of radicals by concentration modulation. By resolving every comb mode of a femtosecond Ti:sapphire laser (3~GHz mode spacing) simultaneously over a wide spectral bandwidth, we have high resolution and absolute frequency accuracy with over 1500 channels measured at once. Figure~\ref{combsetup} provides an overview of the experimental setup, more details can be found in \cite{Sinclair2011}. Briefly, we couple light from the comb into a ring cavity containing an ac discharge cell, which forms the HfF$^+$ and modulates the ions' drift velocity. The resulting modulated Doppler shift produces an absorption signal that is modulated at the discharge frequency. The comb light transmitted through the cavity is then spectrally dispersed using a high-resolution (1~GHz), two-dimensional cross-dispersion system \cite{Adler2010, Diddams2007}, which resolves every comb mode. We then image this onto a lock-in camera \cite{Spirig1997, Beer2005} to demodulate the absorption signal simultaneously on many comb teeth.

We produce HfF$^+$ by heating about 0.5~g of HfF$_4$ powder to about 550~$^\circ$C in a 1~m long home-built discharge cell. Helium buffer gas is flowed through the discharge such that the total pressure is about 6~torr. By recording the emission of the discharge using a low-resolution grating spectrometer, we were able to reliably monitor molecule production. With each loading of the oven we were able to run for about 3 hours. A 2.5~m long bow-tie optical enhancement cavity consisting of two flat, 98\% mirrors (input and output couplers) and two 100~cm radius-of-curvature, low-dispersion, 99.9\% high reflectors surrounds the discharge cell. The reflectivity of the input and output couplers was chosen to match the losses from the Brewster-angled windows and thus provide efficient input coupling. The cavity length is actively stabilized to an integer multiple (25) of the frequency-comb laser cavity length, which ensures that each comb component is coupled to a cavity mode. Our feedback loop provides about 20 kHz of bandwidth by using a high-speed, low-range piezo-electric transducer (PzT)~\cite{Briles2010} and a second long-range PzT. We use liquid-crystal variable retarders and a polarizing beamsplittler to rapidly (50~ms) switch the direction of propagation through the discharge tube while maintaining the comb-cavity lock. This allows us to subtract out slowly varying noise due to drifts in camera pixel offsets and also to improve the rejection of neutral background absorption.

We record about 150~cm$^{-1}$ of spectrum spread over 1500 channels simultaneously. One comb tooth is locked to a stable cw Ti:sapphire laser, which provides an absolute frequency reference in our spectrum. For each measurement, we average and subtract images for each direction of propagation and also record the power per comb tooth by applying a calibrated amplitude modulation to the laser. In order to fully sample the spectrum, we interleave 30 measurements with the cw laser stepped over 3 GHz. This results in a spectrum that covers 150~cm$^{-1}$ sampled every 100~MHz with an absolute frequency accuracy of 30~MHz (set by a rubidium referenced wavemeter). When the wavemeter was not calibrated for a particular measurement, we set the uncertainty at 100 MHz. For strong bands with fully resolved isotope structure, absolute accuracy of the determination of the band origin is limited predominantly not by statistical errors but by the absolute knowledge of the cw Ti:sapphire laser frequency we use as a reference. For our purposes we were satisfied with the 30-100 MHz (0.001 - 0.003~\cm) absolute accuracy of the wavemeter.  This could be readily improved if needed. For analysis purposes we then interpolate the spectrum onto a fixed 0.001~cm$^{-1}$ grid. This allows us to easily average or combine a collection of different scans. One full scan takes about 30 minutes and results in a single-pass fractional absorption sensitivity of $3\times10^{-7}$. Since one scan contains 45000 channels, this equates to a sensitivity of $4\times10^{-8}$ Hz$^{-1/2}$ (spectral element)$^{-1/2}$, which is the sensitivity that a single-frequency laser system would need, in addition to being able to scan 150 cm$^{-1}$ continuously, to match the performance of the comb-vms system. We recorded spectra over 1000~cm$^{-1}$ with both the oven on and off to check for contamination.

\subsection{Single-frequency vms}

The information obtained from the comb scans, combined with theory predictions, allowed us to scan other bands using single-frequency velocity-modulation spectroscopy as sketched in Figure~\ref{cwsetup}. For these measurements we removed the enhancement cavity and counter-propogated beams from the cw Ti:sapphire laser through the discharge tube. We then subtract these two beams using an auto-balanced photodetector \cite{Hobbs1997}. Due to noise from the discharge, we reached a sensitivity of about $5\times10^{-8}$ Hz$^{-1/2}$ with 1 mW on the detector; this noise seemed to be related to optical pickup from the discharge emission, which was reduced using irises, and to non-common-mode amplitude noise from acoustical pickup. Using this technique we were able to find 15 more bands ranging from 9950 to 14600~cm$^{-1}$ without having to scan the full spectral range. We can continuously scan about 0.5~cm$^{-1}$ in 4 minutes, which is at least 30 times slower than the comb-vms system.

\subsection{Coherent Subtraction}

We have also investigated a novel technique for differential detection that relies on coherent interference between the two counter-propogating laser beams instead of subtraction of photocurrents as discussed previously. This is accomplished by coherently splitting and recombining the two beams using one non-polarizing beamsplitter to form a Sagnac-type interferometer containing the discharge cell and cavity if desired (blue inset to Figure~\ref{coherentsub}). The signal is then detected at the destructive interference, or ``dark port", of the interferometer. In the regime where laser intensity noise dominates, the signal-to-noise using coherent subtraction increases as the splitting ratio approaches 50:50 between the two beams until detector readout noise or shot noise dominates. Physically, the fractional signal is increasing while the fractional noise level remains constant and dc power are both decreasing. We have tested coherent subtraction versus a single direction of propagation using a photodiode and single-channel lock-in detector with our cw laser and demonstrate a factor of 10 gain in the signal to noise (blue data in Figure~\ref{coherentsub}), which is the predicted improvement with the 52/48 beamsplitter that was used.

Coherent subtraction, unlike subtraction of photocurrents, is sensitive to phase as well as amplitude of the light field; thus, the resulting lineshape is dependent on both absorption (approximately first-derivative shaped) and dispersion (approximately second-derivative) and can vary if there are differential phase shifts between the two directions. We performed a simple simulation, shown as dashed lines in Fig.~\ref{coherentsub}, that reproduces the lineshape modification. These simulations were done by applying a cosine modulation in time to the center frequency for both the absorption (assumed Gaussian lineshape) and dispersion terms (obtained via Kramers-Kronig) and then selecting the appropriate frequency term from the Fourier cosine transformation. Repeating this at a variety of simulated laser frequencies generates the lineshape. The amplitude and linewidth of the single-direction simulation was scaled to the measured value and these values were used for the coherent-subtraction simulation.  The constant phase offset (most likely due to the beamsplitter) added to the coherent subtraction simulation was chosen to fit the measured lineshape (about $\pi/20$ radians). In addition to the phase sensitivity, which could be useful for some applications, coherent subtraction has several other advantages. First, the dark port can be used to reduce the optical power seen by the detector thus avoiding saturation while still maintaining a shot-noise limited signal-to-noise ratio. Second, the subtraction is performed prior to the detector and is wavelength independent. This means that coherent subtraction can be used with comb-vms or other dispersive detection systems for increased signal-to-noise. We did not implement coherent subtraction with the comb-vms system here because we were close to camera-noise limited, thus the gains would not have been very significant.

\begin{figure*}[p]
\includegraphics[scale=1]{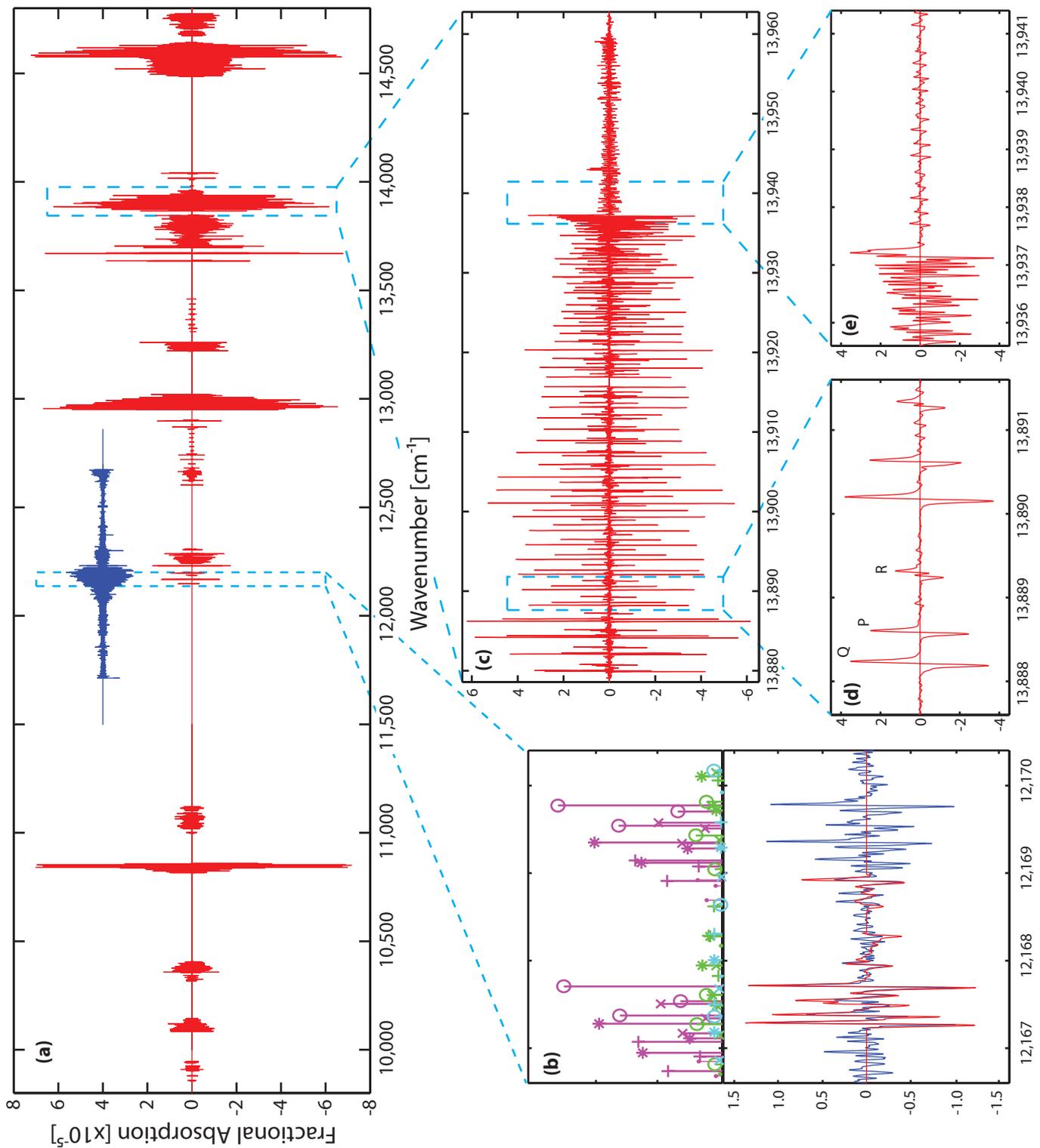}
\caption{\textbf{All data obtained with cw- and comb-vms (red and blue, respectively).} All y-axes are fractional single-pass absorption ($\times10^{-5}$) and all x-axes are in cm$^{-1}$. In (a) we show all of the HfF$^+$ spectra acquired to date. The comb-vms data is vertically offset for clarity. The lower panel of (b) shows a zoomed region of the comb data where some lines were also measured with the cw system. The upper panel of (b) plots predicted line positions from the fits; different colors correspond to different bands (magenta, $^{1}\Pi_{1}\!\leftarrow\!^{1}\Sigma^{+}$ (0,1); green, $^3\Pi_1\!\leftarrow\!^1\Sigma^+$ (3, 1); cyan, $^3\Pi_1\!\leftarrow\!^1\Sigma^+$ (2, 0)) while different symbols correspond to Hf isotopes: $-\circ$, $^{180}$Hf; $-\times$, $^{179}$Hf; $-\ast$, $^{178}$Hf; $-+$, $^{177}$Hf; $-\cdot$, $^{176}$Hf. Inset (c) shows the $^3\Phi_2\!\leftarrow\!^3\Delta_1$ (0, 0) band (on the left) and part of the $^3\Sigma_{0^+}^-\!\leftarrow\!^1\Sigma^+$ (1, 0) on the right. The region in (d) illustrates the cancellation of isotope shifts in $\Delta v = 0$ transitions; the sharpness of P- and Q-branch lines is due to a cancellation between the rotational, vibrational, and electronic contributions to the isotope shift. The splitting in the (higher-$J''$) R-branch lines that is not observed in the (lower-$J''$) P- and Q-branch lines is due to $\Lambda$-doubling in the $^3\Delta_1$ state. Inset (e) shows a prominent band-head from the $^3\Phi_2\!\leftarrow\!^3\Delta_1$ (0, 0) band as well as weaker lines from the $^3\Sigma_{0^+}^-\!\leftarrow\!^1\Sigma^+$ (1, 0) band.}
\label{data}
\end{figure*}

\section{Results and Discussion}
\label{results}
Figure~\ref{data}(a) shows the spectrum of HfF$^+$  acquired using the comb-vms system in blue (offset for clarity) as well as all data obtained with single-frequency vms in red. As illustrated in the zoomed region shown in the lower panel of Figure~\ref{data}(b), the spectrum from about 12100-12300~cm$^{-1}$ is extremely congested due to the presence of many bands, each with five isotopes, and the high temperature of our oven, which results in observed $J''$ values up to about 70. The dynamic range of frequency-comb velocity-modulation spectroscopy is demonstrated by our ability to identify the overlapping $^1\Pi_1\!\leftarrow\!^1\Sigma^+$ ($v' = 0$, $v'' = 1$) and $^{3}\Pi_1\!\leftarrow\!^1\Sigma^+$ (3,1) bands despite the difference in linestrengths and an offset in band origin of only $\sim$1 cm$^{-1}$. Since 150~cm$^{-1}$ sections are acquired simultaneously when using the comb, relative linestrengths within the region are not influenced by variability in oven and discharge conditions, which significantly helps to disentangle bands. The upper panel of Figure~\ref{data}(b) shows predicted line positions from fits to three different bands (the fitting is discussed below), which illustrates our ability to resolve each isotope ($-\circ$, $^{180}$Hf; $-\times$, $^{179}$Hf; $-\ast$, $^{178}$Hf; $-+$, $^{177}$Hf; $-\cdot$, $^{176}$Hf) for multiple bands. Figure~\ref{data}(b) also demonstrates excellent overlap between the comb and cw spectra. 

\begin{figure*}[t]
\includegraphics[scale=1]{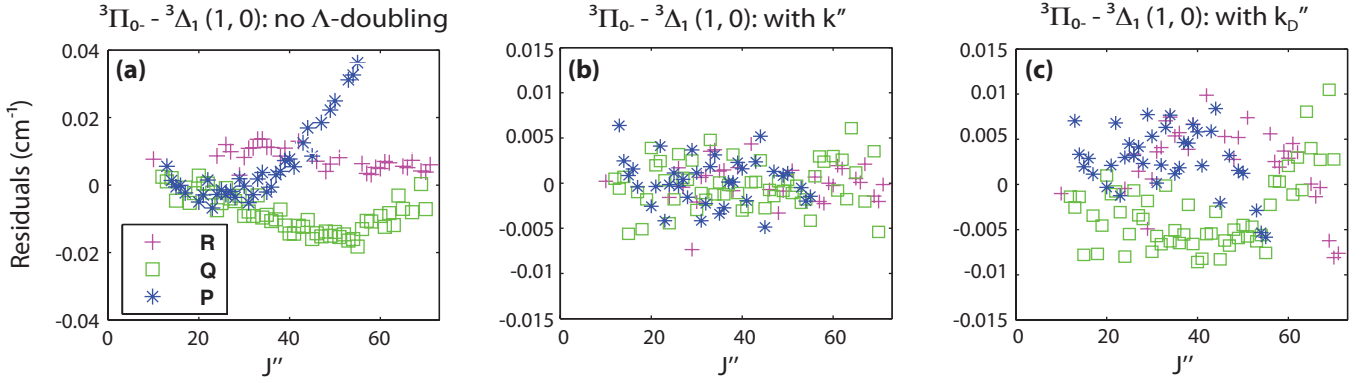}
\caption{\textbf{Residuals for three fits to the $\boldsymbol{^3\Pi_{0^-}\!\leftarrow\!^3\Delta_1}$ (1, 0) band.} The residuals are only plotted for assigned lines in each branch: R-branch (+), Q-branch(square), and P-branch (*). (a) Fit with no $\Lambda$-doubling terms included. Notice the large structure to the residuals. (b) Only $k''$ included, i.e., only $\Lambda$-doubling proportional to $J^2$ in the lower state. No structure in the residuals; the scatter is primarily due to error in identification of the line center. (c) Only $k_D''$ included, i.e., only $\Lambda$-doubling proportional to $J^4$ in the lower state. There is some slight structure to the residuals, indicating that the primary contribution to the $\Lambda$-doubling goes as $J^2$.}
\label{residuals}
\end{figure*}

Since we were fitting a variety of different transitions with various combinations of Hund's cases, we chose to fit each band with the general expression given in Eq.~\ref{fiteq} 
\begin{equation}
\nu(v', v'', J', J'', s', s'') =  \nu_0 + F_{v'}(J', s') - F_{v''}(J'', s'')
\label{fiteq}
\end{equation}
where the rotational energies, $F_v$ are
\begin{equation}
F_v(J, s) = (B_v - s\frac{k_v}{2})J(J+1) - (D_v - s\frac{k_{D,v}}{2})J^2(J+1)^2.
\label{F}
\end{equation}
We include rotation (B), centrifugal distortion (D), $\Lambda$-doubling ($k$), and distortion of the $\Lambda$-doubling ($k_D$) for each vibrational level, $v$; $s$ is an e/f-symmetry term where $s = +1$ for e-symmetry and -1 for f-symmetry. $k$ and $k_D$ are generic constants for the $\Lambda$-doubling since we have fit states with several different values of $\Lambda$ and $\Omega$. In $\Sigma_0$ states, only one symmetry term is chosen and $k$ and $k_D$ are both set to 0. $\nu_0$ includes both the electronic and vibrational energies:
\begin{equation}
\nu_0 = T_e' - T_e'' + E_{vib},
\label{nu_0}
\end{equation}
\begin{multline}
E_{vib} \approx (\omega_e'(v' + 1/2) - \omega_ex_e'(v'+1/2)^2) \\
- (\omega_e''(v'' + 1/2) - \omega_ex_e''(v''+1/2)^2).
\label{Evib}
\end{multline}
Here $T_e$ is defined as the energy difference between the minimum of the potential curve from the minimum of the X$^1\Sigma^+$ curve and $\omega_e$ and $\omega_e x_e$ are the usual vibrational constant and anharmonic correction, respectively.

Due to the congested nature of the spectra, fits were performed to a manually identified list of isolated lines (position only) for each band and for isotope in bands with well resolved isotope structure. Figure~\ref{residuals}(b) shows the residuals from a representative fit; root-mean-square values for the residuals were typically $<0.005$~cm$^{-1}$ with no apparent structure. For most $\Delta v = 0$ transitions the fits were performed to estimated isotope averaged line centers and then the rotational constants were shifted by reduced mass to the $^{180}$Hf values.  The $^3\Phi_2\!\leftarrow\!^3\Delta_1$ bands exhibited line doubling at high-J due to $\Lambda$-doubling (discussed in detail below), for these bands the center of the doublet was used for the fits. Similarly, the $^1\Pi_1\!\leftarrow\!^3\Delta_2$ (0, 0) band exhibited doubling of every transition (i.e., 6 branches) due to large $\Lambda$-doubling in the $^1\Pi_1$; for this band all six branches were fit simultaneously by assuming the splitting to each transition was $k' \times J'(J'+1)$. A summary of the fitted constants for each band is given in Table~\ref{transitions}.

Assignments of the observed bands were not particularly straightforward, partially because the presence or absence of low-J lines usually could not be determined due to the high temperature and complicating isotope structure. In fact, only for the $^1\Pi_1\!\leftarrow\!^1\Sigma^+$ (0, 0) band near 13000~cm$^{-1}$ were we able to directly establish that $\Omega '=1$ and $\Omega '' = 0$ since we observed an R(0) line but no P(1) line. We also measured several bands with no Q-branch, which were assigned as $^3\Pi_{0^+}\!\leftarrow\!^1\Sigma^+$ and $^3\Sigma_{0^+}^-\!\leftarrow\!^1\Sigma^+$. This information, combined with data from \cite{Barker2011}, isotope shifts, and $\Lambda$-doubling structure provided the assignments in Table~\ref{transitions}. Due to possible ambiguities, several bands are still unassigned as indicated.

From the fitted constants for the 16 assigned bands, we can determine constants for the X$^1\Sigma^+$, $^3\Delta_1$, $^3\Pi_{0^-}$, $^3\Pi_{0^+}$, $^1\Pi_1$, $^3\Pi_1$, $^3\Sigma_{0^+}^-$, and $^3\Phi_2$ states as given in Table~\ref{stateconstants}. $T_0$ and $\Delta$G$_{1/2}$ are directly obtained from the measurements: $T_0$ is defined as the energy of the $v=0$ level of an excited state relative to the X$^1\Sigma^+$ $v=0$ level, and $\Delta$G$_{1/2}$ is the energy difference between the $v=0$ and $v=1$ levels. All other (equilibrium) constants are extracted from the data. The rotational constant for a given vibrational level is given to first order in $v$ as $B_v = B_e - \alpha_e(v + 1/2)$, where $B_e$ is the equilibrium rotational constant. Since we have measured at least two vibrational levels for each state, we can determine $\alpha_e \approx B_{v} - B_{v+1}$, and then use this to obtain $B_e$ for each state.  $T_e$, $\omega_e$, and $\omega_e x_e$ are determined using Eqs. \ref{nu_0} and \ref{Evib}. For the $^3\Pi_1$ state, we were able to calculate $\omega_e$ and $\omega_e x_e$ directly; for all other states we assumed a Morse potential so that $\omega_e x_e = \alpha_e^2 \omega_e^2 / 36 B_e^3 + \alpha_e \omega_e / 3B_e + B_e$ (i.e., the Pekeris relationship).

In addition, we have observed isotope shifts of the state origins, called $\delta$T$_e$ in Table~\ref{stateconstants}, that we attribute to Hf electronic isotope shifts (due to the finite nuclear charge radius) \cite{Zimmermann1994, Anastassov1994, Zhao1997}. We define the isotope shift between $^{180}$HfF$^+$ and $^{178}$HfF$^+$ for a given spectral line as 
\begin{multline}
\delta E = E^{180} - E^{178} \\
= (1- \frac{\mu_{180}}{\mu_{178}})E_{rot} + (1 - \sqrt{\frac{\mu_{180}}{\mu_{178}}})E_{vib} + \delta T_e' - \delta T_e''. 
\label{isoeq}
\end{multline}
Here $\mu_{180}$ and $\mu_{178}$ are the reduced masses, $E_{rot}$ is the rotational energy in $^{180}$HfF$^+$ (i.e., offset from the band origin), and $E_{vib}$ is the vibrational energy in $^{180}$HfF$^+$. For bands with well resolved isotope structure (i.e., $\Delta v \not= 0$), $\delta T_e' - \delta T_e''$ was determined either by fitting both $\nu_0^{180}$ and $\nu_0^{178}$ or by measuring $\delta E$ near the origin and then subtracting the vibrational contribution. For  $\Delta v = 0$ bands, we found the frequency where the lines were the sharpest, indicating that $\delta E \approx 0$ (as illustrated by the Q- and P-branch lines in Figure~\ref{data}(d)); this location relative to the band origin is used for $E_{rot}$ in Eq.~\ref{isoeq}, which then gave $\delta T_e' - \delta T_e''$. Values of $\abs{\delta T_e' - \delta T_e''}$ were $\lesssim 0.1$ cm$^{-1}$ for all transitions. To obtain the $\delta$T$_e$ values given in Table~\ref{stateconstants}, we set $\delta T_e''(X^1\Sigma^+) = 0$ and calculated relative shifts of the other states. The simplistic model of Eq.~\ref{isoeq} neglects vibrational-band-specific perturbations or other more complicated level-dependent effects; however, results were consistent across different transitions, which supports the assignment of the effect to electronic states, not individual bands. Except for the X$^1\Sigma^+$ state the isotope shifts are all fairly similar with an average value of about -0.06~\cm (-1.7~GHz), which agrees well with the value for the Hf$^+$ $5d^26s^2\rightarrow5d^26s6p$ of -1.8~GHz \cite{Zhao1997}.

Due to our ability to see high-$J$ lines, we were able to measure precise values for the $\Lambda$-doubling, parametrized by $k$ and $k_D$ in Eq. \ref{F}, in several different electronic levels. In Hund's case (a) $k$ can be related to the more standard $\Lambda$-doubling parameters defined in \cite{Brown1979} for $\Pi$ states ($o$, $p$, and $q$) and to $\tilde{o}_{\Delta}$ as defined in \cite{Brown1987} for $^3\Delta$ states; $k$ was not measurable in $\Phi$ states. The values of $k$ and $k_D$ were useful in assigning many transitions and also provide some insight into inter-state interactions. For the eEDM measurement, the most important $\Lambda$-doubling parameter is the splitting of the $J=1$ levels in the $^3\Delta_1$, as this goes into determining several different potential systematic effects. In a case (a) $^3\Delta_1$ state, the $\Lambda$-doubling is expected to be given by $\pm \tilde{o}_\Delta \times J(J+1)$, where the upper and lower signs refer to the two e/f-symmetry levels \cite{Brown1987}. We can make a rough estimate the value of $\tilde{o}_{\Delta}$ by a simple scaling from that measured in TiO or WC \cite{Wang2012}:
\begin{equation}
\tilde{o}_{\Delta} \approx \frac{\zeta^2(\text{Hf}^{2+}) B^2(\text{HfF}^+)}{\zeta^2(\text{WC/TiO}) B^2(\text{WC/TiO})} \times \tilde{o}_{\Delta}(\text{WC/TiO}).
\label{o}
\end{equation}
This scaling relationship arises because the $\tilde{o}_{\Delta}$ term results from the application of two spin-orbit operators (which scale as the atomic spin-orbit, $\zeta$) and two L-uncoupling operators, which scale as the rotational constant B \cite{Brown1987, Wang2012}; however, it neglects changes in state order and spacing. Using this, with $\zeta$ obtained from \cite{Klinkenberg1961}, we predict $\tilde{o}_{\Delta} \approx 50$~kHz. The $\Lambda$-doubling in the $^3\Delta_1$ can be observed at high-J in the $^3\Phi_2\!\leftarrow\!^3\Delta_1$ (0, 0) band by comparing neighboring P-, Q-, and R-branch lines, as shown in Figure~\ref{data}(d). Since the R-branch lines have the highest $J$, there is noticeable doubling in these lines compared to the P- or Q-branch lines, which indicates that this doubling is not due to isotope splitting. In this band, the lines are not sufficiently well resolved to permit an accurate determination of the $\Lambda$-doubling constant, but by fitting the $^3\Pi_{0^-}\!\leftarrow\!^3\Delta_1$~(1, 0) transition, we were able to precisely determine the sign and size of the $\Lambda$-doubling. As shown in Figure~\ref{residuals}, the best fit was obtained with $k''$ instead of $k_D''$, implying that the $\Lambda$-doubling goes mainly as $J(J+1)$ as expected (adding $k_D''$ to the fit with $k''$ resulted in no improvement). This fit gives a value of $k'' = -1.23(6) \times 10^{-5}$~cm$^{-1}$, which gives $\tilde{o}_{\Delta} = k''/2 = 6.2(3)\times 10^{-6}$~cm$^{-1} = 185(9)$~kHz. The precision on $\tilde{o}_{\Delta}$ is only about an order of magnitude worse than that obtained for TiO using pure rotational spectroscopy \cite{Namiki1998}, illustrating the ability to obtain high-resolution spectroscopic information from broad-bandwidth ro-vibronic spectra.


$\Lambda$-doubling in the $^3\Pi_0$ can arise from interactions with both $^1\Sigma^+$ and $^3\Sigma^+$ states, which result in a splitting of the $\Omega = 0^-$ and $0^+$ levels that is independent of $J$. We measured this splitting to be 189.099(4)~cm$^{-1}$ with the $0^+$ (e-symmetry) higher in energy than $0^-$ (f). To the extent that the X$^1\Sigma^+$ state is a pure $(s\sigma)^2$ configuration, we can assume that this splitting is from a higher-lying $^3\Sigma^+$ state with an $(s\sigma d\sigma)$ configuration. The spin-orbit interaction between the $^3\Sigma_{0^-}^+$ (f) and the $^3\Pi_{0^-}$ can be estimated by 
\begin{equation}
\frac{(\frac{1}{\sqrt{2}} \langle \sigma_1(\beta) \pi(\beta) | \frac{1}{2} \hat{a} l^+ s^- | \sigma_1(\beta) \sigma_2(\alpha) \rangle)^2}{\Delta E} \approx \frac{\frac{3}{4} \zeta^2(5d)}{\Delta E}.
\label{spinorbit}
\end{equation}
We have used $\langle \pi(\beta) | l^+ s^- | \sigma_2(\alpha) \rangle \approx \sqrt{6}$ in the pure-precession model \cite{Brown1979, Mulliken1931}. Using the measured splitting and the atomic spin-orbit coefficient, we estimate $\Delta E \approx 11000$~cm$^{-1}$, which gives an approximate location of the $^3\Sigma_{0^-}^+$ state to be 21200~cm$^{-1}$. This is close to the predicted position of 21694~cm$^{-1}$ from the new calculations (see Section~\ref{theory}). In a Hund's case (a) basis, this splitting corresponds to the more familiar $2(o+p+q)$ \cite{Brown1979}.

The cause of the $\Lambda$-doubling in the $^3\Pi_1$ state is primarily interaction with the $^3\Sigma_{0^-}^+$ state through the $^3\Pi_0$. This type of interaction is characterized by the parameter $p$ in \cite{Brown1979}. With e-symmetry above f in $^3\Pi_0$ and $^3\Pi_0$ located below $^3\Pi_1$, we would expect e above f in $^3\Pi_1$, which corresponds to a negative value of $k$ as observed. The magnitude of $k$ is expected to be about $4 B p / A_\Pi$, where $A_\Pi$ is the spin-orbit parameter for the $^3\Pi$ manifold and
\begin{equation}
p \approx C \langle ^3\Pi_1 | B L^+ | ^3\Sigma_{0^-}^+ \rangle \langle ^3\Sigma_{0^-}^+ | \hat{a} L^- S^+ | ^3\Pi_{0} \rangle / (E_\Pi - E_\Sigma)
\end{equation}
\begin{equation}
\approx C B \zeta(5d) \langle \pi^+ | l^+ | \sigma_2 \rangle^2 / (E_\Pi - E_\Sigma).
\end{equation}
Here, C is a numerical factor dependent on the spin \cite{Brown1979}. Again using the pure-precession model, we estimate $k \approx 8\times 10^{-4}$~cm$^{-1}$, in reasonable agreement with experiment.

We attribute most of the $\Lambda$-doubling in the $^1\Pi_1$ to interaction with the nearly-degenerate nominal $^3\Sigma_{0^+}^-$ state. It is reasonable that the $^1\Pi$ and $^3\Sigma^-$ are not pure ($s\sigma d \pi$) and $(d \delta)^2$ configurations, which would lead to increased spin-orbit interaction between $^1\Pi_1$ and $^3\Sigma_{0^+}^-$. Estimating the magnitude of the $\Lambda$-doubling in this case is difficult due to uncertainty in the purity of the configurations. Qualitatively though, we would expect the effect of a $^3\Sigma^-$ state above the $^1\Pi_1$ state would be to push e-symmetry below f as observed. In addition we observed $\Lambda$-doubling that was strongly vibrational-level dependent ($3.69\times10^{-4}$~cm$^{-1}$ for the $v=0$ compared with $2.68\times10^{-4}$~cm$^{-1}$ for the $v=1$), which is indicative of nearly-degenerate interacting states. We also measured a $\Lambda$-doubling term proportional to $J^4$ (denoted $k_D$). This term can be explained by substituting $\Delta E \rightarrow \Delta E + (B_\Pi - B_\Sigma) J (J+1)$ in the $\Lambda$-doubling denominator; the first two terms in the Taylor expansion give the $J(J+1)$ and the $J^2 (J+1)^2$ components. From this we can estimate that $k_D \approx 3 k (B_\Pi - B_\Sigma) / (E_\Pi - E_\Sigma)$, which gives $k_D \approx 2\times10^{-8}$~\cm.

 \section{Theory}
 \label{theory}

In earlier theoretical study of HfF$^+$ \cite{Petrov2009}, the correlation calculations of the spectroscopic constants were performed in two different ways:  in the first series 10 electrons from $5d,6s$ shells of Hf and $2s,2p$ of F were correlated while in the second series $5s,5p$ outer core electrons of Hf and $1s$ of F were also correlated.  The rest of the $1s-4d$ inner-core electrons of Hf were excluded from the explicit treatment using the generalized relativistic effective core potential (GRECP) method\footnote{60-electron core GRECP for Hf is available at www.qchem.pnpi.spb.ru} \cite{Mosyagin:10a}.  While it was shown that the inclusion of the additional electrons in the 20-electrons case contributed significantly to the excitation energies, such 20-electron calculations were performed only for one internuclear Hf--F distance in a small space of many-electron basis functions due to computer limitations. This calculation was then applied for other distances as a ``core correction''. In the present work a new 20-electron relativistic correlation calculation was carried out in order to consider electronic states that were not investigated in \cite{Petrov2009} as well as to overcome the bottlenecks of the previous studies of HfF$^+$.  Eventually 23 electronic states (with excitation energies up to 22000~\cm) were considered in the present study instead of ten states in \cite{Petrov2009}.

Some modifications were made in the computational procedure to achieve better accuracy.  We used a direct multi-reference configuration-interaction approach accounting for spin-orbit effects (SODCI) \cite{Alekseyev:04a,Titov:01} as a method to treat both correlation and relativistic effects simultaneously.  One-component basis functions (orbitals) are required in this method to construct many-electron spin-adapted functions (SAF's).  Instead of the orbitals obtained within the complete active space self-consistent field method (used in \cite{Petrov2009}), the natural orbitals of a one-electron density matrix averaged over the density matrices of the states of interest were used. The latter density matrices were calculated within the scalar-relativistic coupled-clusters method with single and double cluster amplitudes (CCSD) using the {\sc cfour} code \cite{CFOUR}.  From numerical investigation \cite{Shavitt:77} it is known that the use of natural orbitals can provide faster convergence in terms of the number of SAF's required to account for a given part of correlation energy.

In the SODCI calculation the same atomic basis sets for Hf and F were used as in our previous study \cite{Petrov2009}, i.e., a generally-contracted basis set for Hf consisting of 6 s-, 5 p-, 5 d-, 3 f- and 1 g- type contracted functions (denoted as \{6,5,5,3,1\}) and an ANO-I \{4,3,2,1\} basis set for F \citep{Roos:05}.  However, a correction on the extension of the basis set was
applied additionally in the present work. To evaluate the correction, two scalar-relativistic calculations were performed using the coupled clusters method with single, double and perturbative triple cluster amplitudes:
(i) in the same basis set that was used at the SODCI stage, and (ii) with an extended basis set for Hf (produced by
     uncontracting
d- and adding g-, h-, i- type basis functions) and an ANO-L \{7,7,4,3\} \citep{Roos:05} basis for F.

To compute potential curves the described calculations were performed at 14 points in the range of 3.0 -- 4.2 a.u.  As shown in Table~\ref{theorycomp}, the new calculations are in remarkable agreement with all of the experimental results. These improvements should enable more accurate calculations in other species with heavy atoms, where relativistic effects are extremely important.

\section{Outlook and Conclusions}
\label{conclusions}
\begin{figure}
\begin{center}
\includegraphics[scale=0.95]{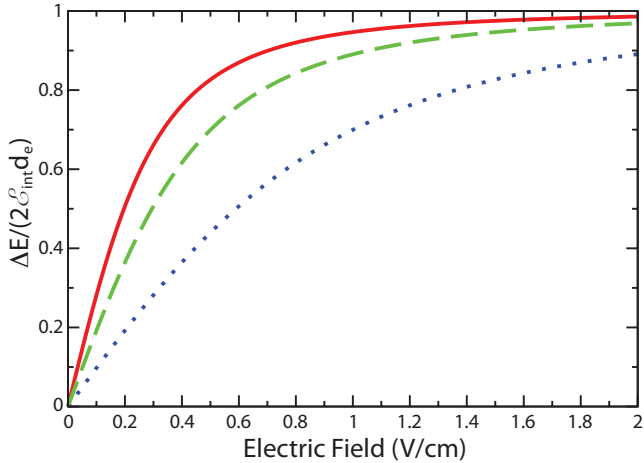}
\caption{\label{polar} 
The eEDM induced Stark splitting ($\Delta E$) for $J=1$ hyperfine levels
of the $^3\Delta_1$ state of HfF$^+$:
between $F=3/2$ $m_F=\pm3/2$ levels (solid line),
between $F=3/2$ $m_F=\pm1/2$ levels (doted line), and
between $F=1/2$ $m_F=\pm1/2$ (dashed line).}
\end{center}
\end{figure}

As shown in Figure~\ref{energylevels} and discussed in the introduction, several different optical transitions may be necessary for the eEDM experiment. We have measured suitable candidates for all of these: (a) can be accomplished using the $^3\Pi_{0^+}$ intermediate, (b) could use the $^3\Phi_2$ level, and (c) could use one of the $^3\Pi_0$ levels or the $^3\Phi_2$ level. Another important value for the eEDM measurements is the size of the $\Lambda$-doubling in the $^3\Delta_1$ $J=1$ level; it is the opposite parity levels in this state that are mixed in an electric field to polarize the molecule. The necessary electric field for full polarization, $\mathcal{E}_{pol}$, is approximately where the Stark energy is larger than the energy difference between the two parity states: thus $\mathcal{E}_{pol} \approx \omega_{ef}/2\pi d_{mf}$ \cite{Meyer2006}, where $\omega_{ef}$ is the $\Lambda$-doubling splitting (in angular frequency) and $d_{mf}$ is the molecular-frame electric dipole moment of the molecule (4.3 Debye for HfF$^+$ \cite{Leanhardt2011}). We can use our measurement of $\tilde{o}_{\Delta}$ to estimate that $\omega_{ef} = 2\pi \times 4 \tilde{o}_{\Delta} \approx 2\pi \times 740$~kHz, which is at least an order magnitude larger than previously predicted \cite{Leanhardt2011}. Nonetheless, this means that $\mathcal{E}_{pol}$ is still under 1~V/cm, which is important for the eEDM experiment as larger fields tend to lead to more issues with systematic errors. As can be seen in Figure~\ref{polar}, numerical results for $^3\Delta_1$ $J=1$ hyperfine levels confirm that at 1~V/cm the splitting in the $\abs{m_F} = 3/2$ levels due to eEDM related Stark shift is almost saturated at $2 d_e \mathcal{E}_{int}$. This saturation occurs as the molecule becomes fully polarized. These results were obtained as described in \cite{Petrov:11}.

We can also use the improved theory calculations to better estimate the radiative lifetime, another important consideration for the eEDM experiment. Using the measured energy separation and calculated dipole moments, we estimate the lifetime of $^3\Delta_1$ $v = 0$ to be about 2 s \cite{Petrov2009}. This is encouraging, as it sets the ultimate limit on the coherence time achievable in the experiment. The high precision that we obtain for rotational constants should enable microwave spectroscopy with minimal searching, which will then be able to resolve the fine and hyperfine structure in the $^3\Delta_1$ level. These measurements are important for estimating systematics and for checking the accuracy of the calculated effective electric field that the electron experiences ($\mathcal{E}_{int}$). We predict that the $^1\Sigma^+(v=0)$ $J'=1 \!\leftarrow\! J''=0$ transition in $^{180}$HfF$^{+}$ will be at 18.290(2) GHz and the $^3\Delta_1(v=0)$ $J'=2 \!\leftarrow\! J''=1$ transition doublet will be at centered at 35.869(2) GHz. To get these frequencies, we used the best measurements of $B_e$ and $\omega_e$ from Table~\ref{theorycomp} to calculate $D_e = 4 B_e^3/\omega_e^2$, then used the measured $B_0$ and calculated $D_e$ to determine the microwave frequencies.

We are currently working on improving the comb-vms system and applying it towards characterization of ThF$^+$, which has several advantages over HfF$^+$ for the eEDM experiment. However for this species, the \textit{ab initio} calculations are even more challenging, leading to larger theoretical uncertainties. The first spectroscopy of ThF$^+$ using PFI-ZEKE plus some LIF (unassigned) has recently been published \cite{Barker2012}, but, as was the case with HfF$^+$, the higher excited states remain uncharacterized. By using highly non-linear fiber to broaden the comb spectrum \cite{Dudley2006} and a high repetition rate Ti:sapphire amplifier \cite{Paul2008}, we hope to cover over 2000~cm$^{-1}$ with comb-vms. This ability to rapidly cover thousands of cm$^{-1}$ with high sensitivity and high resolution will be a powerful new tool for the study of ions and radicals for many applications. In addition, applications to other many other spectral regions are possible by using different comb sources combined with non-linear optics for covering the near-IR \cite{Tauser2004, Cossel2010} or mid-IR \cite{Sun2007, Gambetta2008, Adler2009, Leindecker2011, Neely2011}. These sources plus readout systems using different cameras or optical up-conversion \cite{Johnson2011a} create the possibility of ion spectroscopy anywhere from the visible to the mid-IR.

\textit{Acknowledgements.} Funding for the work at JILA was provided by the NSF, NIST and the Marsico Foundation. We thank R. Stutz, H. Loh, F. Adler, and M.J. Thorpe for many useful discussions. RWF acknowledges support from NSF grant CHE-1058709. Funding for the work at PNPI was provided by Russian Ministry of Education and Science, contract \#07.514.11.4141 (2012-2013). L.S. is grateful also to the Dmitry Zimin ``Dynasty'' Foundation. The molecular calculations were performed at the Supercomputer ``Lomonosov''.




\bibliographystyle{model1a-num-names}
\bibliography{library}

\begin{thebibliography}{72}
\expandafter\ifx\csname natexlab\endcsname\relax\def\natexlab#1{#1}\fi
\providecommand{\bibinfo}[2]{#2}
\ifx\xfnm\relax \def\xfnm[#1]{\unskip,\space#1}\fi
\bibitem[{Hinds(1997)}]{Hinds1997}
\bibinfo{author}{E.~A. Hinds}, \bibinfo{journal}{Physica Scripta}
  \bibinfo{volume}{T70} (\bibinfo{year}{1997}) \bibinfo{pages}{34--41}.
\bibitem[{Sandars(2001)}]{Sandars2001}
\bibinfo{author}{P.~G.~H. Sandars}, \bibinfo{journal}{Contemporary Physics}
  \bibinfo{volume}{42} (\bibinfo{year}{2001}) \bibinfo{pages}{97--111}.
\bibitem[{Fortson et~al.(2003)Fortson, Sandars, and Barr}]{Fortson2003}
\bibinfo{author}{N.~Fortson}, \bibinfo{author}{P.~Sandars},
  \bibinfo{author}{S.~Barr}, \bibinfo{journal}{Physics Today}
  (\bibinfo{year}{2003}) \bibinfo{pages}{33--39}.
\bibitem[{Regan et~al.(2002)Regan, Commins, Schmidt, and DeMille}]{Regan2002}
\bibinfo{author}{B.~C. Regan}, \bibinfo{author}{E.~D. Commins},
  \bibinfo{author}{C.~J. Schmidt}, \bibinfo{author}{D.~DeMille},
  \bibinfo{journal}{Physical Review Letters} \bibinfo{volume}{88}
  (\bibinfo{year}{2002}) \bibinfo{pages}{18--21}.
\bibitem[{Hudson et~al.(2011)Hudson, Kara, Smallman, Sauer, Tarbutt, and
  Hinds}]{Hudson2011}
\bibinfo{author}{J.~J. Hudson}, \bibinfo{author}{D.~M. Kara},
  \bibinfo{author}{I.~J. Smallman}, \bibinfo{author}{B.~E. Sauer},
  \bibinfo{author}{M.~R. Tarbutt}, \bibinfo{author}{E.~A. Hinds},
  \bibinfo{journal}{Nature} \bibinfo{volume}{473} (\bibinfo{year}{2011})
  \bibinfo{pages}{493--6}.
\bibitem[{Commins(1999)}]{Commins1999}
\bibinfo{author}{E.~D. Commins}, \bibinfo{journal}{Advances In Atomic,
  Molecular, and Optical Physics} \bibinfo{volume}{40} (\bibinfo{year}{1999})
  \bibinfo{pages}{1--55}.
\bibitem[{Meyer et~al.(2006)Meyer, Bohn, and Deskevich}]{Meyer2006}
\bibinfo{author}{E.~R. Meyer}, \bibinfo{author}{J.~L. Bohn},
  \bibinfo{author}{M.~Deskevich}, \bibinfo{journal}{Physical Review A}
  \bibinfo{volume}{73} (\bibinfo{year}{2006}) \bibinfo{pages}{1--10}.
\bibitem[{Meyer and Bohn(2008)}]{Meyer2008}
\bibinfo{author}{E.~R. Meyer}, \bibinfo{author}{J.~L. Bohn},
  \bibinfo{journal}{Physical Review A} \bibinfo{volume}{78}
  (\bibinfo{year}{2008}) \bibinfo{pages}{010502(R)}.
\bibitem[{Petrov et~al.(2007)Petrov, Mosyagin, Isaev, and Titov}]{Petrov2007}
\bibinfo{author}{A.~N. Petrov}, \bibinfo{author}{N.~S. Mosyagin},
  \bibinfo{author}{T.~A. Isaev}, \bibinfo{author}{A.~V. Titov},
  \bibinfo{journal}{Physical Review A} \bibinfo{volume}{76}
  (\bibinfo{year}{2007}) \bibinfo{pages}{3--6}.
\bibitem[{Leanhardt et~al.(2011)Leanhardt, Bohn, Loh, Maletinsky, Meyer,
  Sinclair, Stutz, and Cornell}]{Leanhardt2011}
\bibinfo{author}{A.~E. Leanhardt}, \bibinfo{author}{J.~L. Bohn},
  \bibinfo{author}{H.~Loh}, \bibinfo{author}{P.~Maletinsky},
  \bibinfo{author}{E.~R. Meyer}, \bibinfo{author}{L.~C. Sinclair},
  \bibinfo{author}{R.~P. Stutz}, \bibinfo{author}{E.~A. Cornell},
  \bibinfo{journal}{Journal of Molecular Spectroscopy} \bibinfo{volume}{270}
  (\bibinfo{year}{2011}) \bibinfo{pages}{1--25}.
\bibitem[{DeMille et~al.(2001)DeMille, Bay, Bickman, Kawall, Hunter, {Krause,
  Jr.}, Maxwell, and Ulmer}]{DeMille2001}
\bibinfo{author}{D.~DeMille}, \bibinfo{author}{F.~Bay},
  \bibinfo{author}{S.~Bickman}, \bibinfo{author}{D.~Kawall},
  \bibinfo{author}{L.~Hunter}, \bibinfo{author}{D.~{Krause, Jr.}},
  \bibinfo{author}{S.~Maxwell}, \bibinfo{author}{K.~Ulmer},
  \bibinfo{journal}{AIP Conference Proceedings} \bibinfo{volume}{596}
  (\bibinfo{year}{2001}) \bibinfo{pages}{72--83}.
\bibitem[{Barker et~al.(2011)Barker, Antonov, Bondybey, and
  Heaven}]{Barker2011}
\bibinfo{author}{B.~J. Barker}, \bibinfo{author}{I.~O. Antonov},
  \bibinfo{author}{V.~E. Bondybey}, \bibinfo{author}{M.~C. Heaven},
  \bibinfo{journal}{The Journal of Chemical Physics} \bibinfo{volume}{134}
  (\bibinfo{year}{2011}) \bibinfo{pages}{201102}.
\bibitem[{Barker et~al.(2012)Barker, Antonov, Heaven, and
  Peterson}]{Barker2012}
\bibinfo{author}{B.~J. Barker}, \bibinfo{author}{I.~O. Antonov},
  \bibinfo{author}{M.~C. Heaven}, \bibinfo{author}{K.~A. Peterson},
  \bibinfo{journal}{The Journal of Chemical Physics} \bibinfo{volume}{136}
  (\bibinfo{year}{2012}) \bibinfo{pages}{104305}.
\bibitem[{Kaledin et~al.(1995)Kaledin, Mccord, and Heaven}]{Kaledin1995}
\bibinfo{author}{L.~Kaledin}, \bibinfo{author}{J.~Mccord},
  \bibinfo{author}{M.~Heaven}, \bibinfo{journal}{Journal of Molecular
  Spectroscopy} \bibinfo{volume}{173} (\bibinfo{year}{1995})
  \bibinfo{pages}{499--509}.
\bibitem[{Namiki et~al.(1998)Namiki, Saito, Robinson, and Steimle}]{Namiki1998}
\bibinfo{author}{K.~Namiki}, \bibinfo{author}{S.~Saito}, \bibinfo{author}{J.~S.
  Robinson}, \bibinfo{author}{T.~C. Steimle}, \bibinfo{journal}{Journal of
  Molecular Spectroscopy} \bibinfo{volume}{191} (\bibinfo{year}{1998})
  \bibinfo{pages}{176--182}.
\bibitem[{Kobayashi et~al.(2002)Kobayashi, Hall, Muckerman, Sears, and
  Merer}]{Kobayashi2002}
\bibinfo{author}{K.~Kobayashi}, \bibinfo{author}{G.~E. Hall},
  \bibinfo{author}{J.~T. Muckerman}, \bibinfo{author}{T.~J. Sears},
  \bibinfo{author}{A.~J. Merer}, \bibinfo{journal}{Journal of Molecular
  Spectroscopy} \bibinfo{volume}{212} (\bibinfo{year}{2002})
  \bibinfo{pages}{133--141}.
\bibitem[{Focsa et~al.(1998)Focsa, Pinchemel, Collet, and Huet}]{Focsa1998}
\bibinfo{author}{C.~Focsa}, \bibinfo{author}{B.~Pinchemel},
  \bibinfo{author}{D.~Collet}, \bibinfo{author}{T.~Huet},
  \bibinfo{journal}{Journal of Molecular Spectroscopy} \bibinfo{volume}{189}
  (\bibinfo{year}{1998}) \bibinfo{pages}{254--263}.
\bibitem[{Focsa and Pinchemel(1999)}]{Focsa1999}
\bibinfo{author}{C.~Focsa}, \bibinfo{author}{B.~Pinchemel},
  \bibinfo{journal}{Chemical Physics} \bibinfo{volume}{247}
  (\bibinfo{year}{1999}) \bibinfo{pages}{395--405}.
\bibitem[{Petrov et~al.(2009)Petrov, Mosyagin, and Titov}]{Petrov2009}
\bibinfo{author}{A.~N. Petrov}, \bibinfo{author}{N.~S. Mosyagin},
  \bibinfo{author}{A.~V. Titov}, \bibinfo{journal}{Physical Review A}
  \bibinfo{volume}{79} (\bibinfo{year}{2009}) \bibinfo{pages}{1--7}.
\bibitem[{Sinclair et~al.(2011)Sinclair, Cossel, Coffey, Ye, and
  Cornell}]{Sinclair2011}
\bibinfo{author}{L.~C. Sinclair}, \bibinfo{author}{K.~C. Cossel},
  \bibinfo{author}{T.~Coffey}, \bibinfo{author}{J.~Ye}, \bibinfo{author}{E.~A.
  Cornell}, \bibinfo{journal}{Physical Review Letters} \bibinfo{volume}{107}
  (\bibinfo{year}{2011}) \bibinfo{pages}{093002}.
\bibitem[{Klemperer(2011)}]{Klemperer2011}
\bibinfo{author}{W.~Klemperer}, \bibinfo{journal}{Annual Review of Physical
  Chemistry} \bibinfo{volume}{62} (\bibinfo{year}{2011})
  \bibinfo{pages}{173--184}.
\bibitem[{Snow and Bierbaum(2008)}]{Snow2008}
\bibinfo{author}{T.~P. Snow}, \bibinfo{author}{V.~M. Bierbaum},
  \bibinfo{journal}{Annual Review of Analytical Chemistry} \bibinfo{volume}{1}
  (\bibinfo{year}{2008}) \bibinfo{pages}{229--59}.
\bibitem[{Indriolo et~al.(2010)Indriolo, Oka, Geballe, and
  McCall}]{Indriolo2010}
\bibinfo{author}{N.~Indriolo}, \bibinfo{author}{T.~Oka}, \bibinfo{author}{T.~R.
  Geballe}, \bibinfo{author}{B.~J. McCall}, \bibinfo{journal}{The Astrophysical
  Journal} \bibinfo{volume}{711} (\bibinfo{year}{2010})
  \bibinfo{pages}{1338--1342}.
\bibitem[{Sarre(2006)}]{Sarre2006}
\bibinfo{author}{P.~J. Sarre}, \bibinfo{journal}{Journal of Molecular
  Spectroscopy} \bibinfo{volume}{238} (\bibinfo{year}{2006})
  \bibinfo{pages}{1--10}.
\bibitem[{Oka and McCall(2011)}]{Oka2011}
\bibinfo{author}{T.~Oka}, \bibinfo{author}{B.~J. McCall},
  \bibinfo{journal}{Science (New York, N.Y.)} \bibinfo{volume}{331}
  (\bibinfo{year}{2011}) \bibinfo{pages}{293--4}.
\bibitem[{Snow and McCall(2006)}]{Snow2006}
\bibinfo{author}{T.~P. Snow}, \bibinfo{author}{B.~J. McCall},
  \bibinfo{journal}{Annual Review of Astronomy and Astrophysics}
  \bibinfo{volume}{44} (\bibinfo{year}{2006}) \bibinfo{pages}{367--414}.
\bibitem[{Halfen and Ziurys(2007)}]{Halfen2007}
\bibinfo{author}{D.~T. Halfen}, \bibinfo{author}{L.~M. Ziurys},
  \bibinfo{journal}{The Astrophysical Journal} \bibinfo{volume}{657}
  (\bibinfo{year}{2007}) \bibinfo{pages}{L61--L64}.
\bibitem[{Maier et~al.(2011)Maier, Walker, Bohlender, Mazzotti, Raghunandan,
  Fulara, Garkusha, and Nagy}]{Maier2011}
\bibinfo{author}{J.~P. Maier}, \bibinfo{author}{G.~A.~H. Walker},
  \bibinfo{author}{D.~A. Bohlender}, \bibinfo{author}{F.~J. Mazzotti},
  \bibinfo{author}{R.~Raghunandan}, \bibinfo{author}{J.~Fulara},
  \bibinfo{author}{I.~Garkusha}, \bibinfo{author}{A.~Nagy},
  \bibinfo{journal}{The Astrophysical Journal} \bibinfo{volume}{726}
  (\bibinfo{year}{2011}) \bibinfo{pages}{41}.
\bibitem[{Gottfried et~al.(2003)Gottfried, McCall, and Oka}]{Gottfried2003}
\bibinfo{author}{J.~L. Gottfried}, \bibinfo{author}{B.~J. McCall},
  \bibinfo{author}{T.~Oka}, \bibinfo{journal}{The Journal of Chemical Physics}
  \bibinfo{volume}{118} (\bibinfo{year}{2003}) \bibinfo{pages}{10890}.
\bibitem[{Morong et~al.(2009)Morong, Gottfried, and Oka}]{Morong2009}
\bibinfo{author}{C.~P. Morong}, \bibinfo{author}{J.~L. Gottfried},
  \bibinfo{author}{T.~Oka}, \bibinfo{journal}{Journal of Molecular
  Spectroscopy} \bibinfo{volume}{255} (\bibinfo{year}{2009})
  \bibinfo{pages}{13--23}.
\bibitem[{Pavanello et~al.(2012)Pavanello, Adamowicz, Alijah, Zobov, Mizus,
  Polyansky, Tennyson, Szidarovszky, Cs\'{a}sz\'{a}r, Berg, Petrignani, and
  Wolf}]{Pavanello2012}
\bibinfo{author}{M.~Pavanello}, \bibinfo{author}{L.~Adamowicz},
  \bibinfo{author}{A.~Alijah}, \bibinfo{author}{N.~Zobov},
  \bibinfo{author}{I.~Mizus}, \bibinfo{author}{O.~L. Polyansky},
  \bibinfo{author}{J.~Tennyson}, \bibinfo{author}{T.~Szidarovszky},
  \bibinfo{author}{A.~G. Cs\'{a}sz\'{a}r}, \bibinfo{author}{M.~Berg},
  \bibinfo{author}{A.~Petrignani}, \bibinfo{author}{A.~Wolf},
  \bibinfo{journal}{Physical Review Letters} \bibinfo{volume}{108}
  (\bibinfo{year}{2012}) \bibinfo{pages}{023002}.
\bibitem[{White(1999)}]{White1999}
\bibinfo{author}{E.~T. White}, \bibinfo{journal}{Science} \bibinfo{volume}{284}
  (\bibinfo{year}{1999}) \bibinfo{pages}{135--137}.
\bibitem[{Asvany et~al.(2005)Asvany, P, Redlich, Hegemann, Schlemmer, and
  Marx}]{Asvany2005}
\bibinfo{author}{O.~Asvany}, \bibinfo{author}{P.~K. P},
  \bibinfo{author}{B.~Redlich}, \bibinfo{author}{I.~Hegemann},
  \bibinfo{author}{S.~Schlemmer}, \bibinfo{author}{D.~Marx},
  \bibinfo{journal}{Science} \bibinfo{volume}{309} (\bibinfo{year}{2005})
  \bibinfo{pages}{1219--22}.
\bibitem[{Huang et~al.(2006)Huang, McCoy, Bowman, Johnson, Savage, Dong, and
  Nesbitt}]{Huang2006}
\bibinfo{author}{X.~Huang}, \bibinfo{author}{A.~B. McCoy},
  \bibinfo{author}{J.~M. Bowman}, \bibinfo{author}{L.~M. Johnson},
  \bibinfo{author}{C.~Savage}, \bibinfo{author}{F.~Dong},
  \bibinfo{author}{D.~J. Nesbitt}, \bibinfo{journal}{Science}
  \bibinfo{volume}{311} (\bibinfo{year}{2006}) \bibinfo{pages}{60--3}.
\bibitem[{Mills et~al.(2011)Mills, Siller, Porambo, Perera, Kreckel, and
  McCall}]{Mills2011}
\bibinfo{author}{A.~A. Mills}, \bibinfo{author}{B.~M. Siller},
  \bibinfo{author}{M.~W. Porambo}, \bibinfo{author}{M.~Perera},
  \bibinfo{author}{H.~Kreckel}, \bibinfo{author}{B.~J. McCall},
  \bibinfo{journal}{The Journal of Chemical Physics} \bibinfo{volume}{135}
  (\bibinfo{year}{2011}) \bibinfo{pages}{224201}.
\bibitem[{Hobbs(1997)}]{Hobbs1997}
\bibinfo{author}{P.~C. Hobbs}, \bibinfo{journal}{Applied Optics}
  \bibinfo{volume}{36} (\bibinfo{year}{1997}) \bibinfo{pages}{903--20}.
\bibitem[{Thorpe et~al.(2006)Thorpe, Moll, Jones, Safdi, and Ye}]{Thorpe2006}
\bibinfo{author}{M.~J. Thorpe}, \bibinfo{author}{K.~D. Moll},
  \bibinfo{author}{R.~J. Jones}, \bibinfo{author}{B.~Safdi},
  \bibinfo{author}{J.~Ye}, \bibinfo{journal}{Science} \bibinfo{volume}{311}
  (\bibinfo{year}{2006}) \bibinfo{pages}{1595--9}.
\bibitem[{Thorpe and Ye(2008)}]{Thorpe2008}
\bibinfo{author}{M.~J. Thorpe}, \bibinfo{author}{J.~Ye},
  \bibinfo{journal}{Applied Physics B} \bibinfo{volume}{91}
  (\bibinfo{year}{2008}) \bibinfo{pages}{397--414}.
\bibitem[{Adler et~al.(2010)Adler, Thorpe, Cossel, and Ye}]{Adler2010}
\bibinfo{author}{F.~Adler}, \bibinfo{author}{M.~J. Thorpe},
  \bibinfo{author}{K.~C. Cossel}, \bibinfo{author}{J.~Ye},
  \bibinfo{journal}{Annual Review of Analytical Chemistry} \bibinfo{volume}{3}
  (\bibinfo{year}{2010}) \bibinfo{pages}{175--205}.
\bibitem[{Gudeman et~al.(1983)Gudeman, Begemann, Pfaff, and
  Saykally}]{Gudeman1983}
\bibinfo{author}{C.~Gudeman}, \bibinfo{author}{M.~Begemann},
  \bibinfo{author}{J.~Pfaff}, \bibinfo{author}{R.~J. Saykally},
  \bibinfo{journal}{Physical Review Letters} \bibinfo{volume}{50}
  (\bibinfo{year}{1983}) \bibinfo{pages}{727--731}.
\bibitem[{Saykally(1988)}]{Saykally1988a}
\bibinfo{author}{R.~J. Saykally}, \bibinfo{journal}{Science}
  \bibinfo{volume}{239} (\bibinfo{year}{1988}) \bibinfo{pages}{157--61}.
\bibitem[{Lindsay et~al.(2001)Lindsay, {Rade Jr.}, and Oka}]{Lindsay2001}
\bibinfo{author}{C.~Lindsay}, \bibinfo{author}{R.~M. {Rade Jr.}},
  \bibinfo{author}{T.~Oka}, \bibinfo{journal}{Journal of Molecular
  Spectroscopy} \bibinfo{volume}{210} (\bibinfo{year}{2001})
  \bibinfo{pages}{51--59}.
\bibitem[{Stephenson and Saykally(2005)}]{Stephenson2005}
\bibinfo{author}{S.~K. Stephenson}, \bibinfo{author}{R.~J. Saykally},
  \bibinfo{journal}{Chemical Reviews} \bibinfo{volume}{105}
  (\bibinfo{year}{2005}) \bibinfo{pages}{3220--34}.
\bibitem[{Diddams et~al.(2007)Diddams, Hollberg, and Mbele}]{Diddams2007}
\bibinfo{author}{S.~A. Diddams}, \bibinfo{author}{L.~Hollberg},
  \bibinfo{author}{V.~Mbele}, \bibinfo{journal}{Nature} \bibinfo{volume}{445}
  (\bibinfo{year}{2007}) \bibinfo{pages}{627--30}.
\bibitem[{Spirig et~al.(1997)Spirig, Marley, and Seitz}]{Spirig1997}
\bibinfo{author}{T.~Spirig}, \bibinfo{author}{M.~Marley},
  \bibinfo{author}{P.~Seitz}, \bibinfo{journal}{IEEE Transactions on Electron
  Devices} \bibinfo{volume}{44} (\bibinfo{year}{1997})
  \bibinfo{pages}{1643--1647}.
\bibitem[{Beer and Seitz(2005)}]{Beer2005}
\bibinfo{author}{S.~Beer}, \bibinfo{author}{P.~Seitz},
  \bibinfo{journal}{Research in Microelectronics and Electronics, 2005 PhD}
  \bibinfo{volume}{2} (\bibinfo{year}{2005}) \bibinfo{pages}{135--138}.
\bibitem[{Briles et~al.(2010)Briles, Yost, Cing\"{o}z, Ye, and
  Schibli}]{Briles2010}
\bibinfo{author}{T.~C. Briles}, \bibinfo{author}{D.~C. Yost},
  \bibinfo{author}{A.~Cing\"{o}z}, \bibinfo{author}{J.~Ye},
  \bibinfo{author}{T.~R. Schibli}, \bibinfo{journal}{Optics Express}
  \bibinfo{volume}{18} (\bibinfo{year}{2010}) \bibinfo{pages}{9739--46}.
\bibitem[{Zimmermann et~al.(1994)Zimmermann, Baumann, Kuszner, and
  Werner}]{Zimmermann1994}
\bibinfo{author}{D.~Zimmermann}, \bibinfo{author}{P.~Baumann},
  \bibinfo{author}{D.~Kuszner}, \bibinfo{author}{A.~Werner},
  \bibinfo{journal}{Physical Review A} \bibinfo{volume}{50}
  (\bibinfo{year}{1994}) \bibinfo{pages}{1112--1120}.
\bibitem[{Anastassov et~al.(1994)Anastassov, Gangrsky, Kul'djanov, Marinova,
  Markov, and Zemlyanoi}]{Anastassov1994}
\bibinfo{author}{A.~Anastassov}, \bibinfo{author}{Y.~Gangrsky},
  \bibinfo{author}{B.~K. Kul'djanov}, \bibinfo{author}{K.~P. Marinova},
  \bibinfo{author}{B.~N. Markov}, \bibinfo{author}{S.~G. Zemlyanoi},
  \bibinfo{journal}{Zeitschrift f\"{u}r Physik A} \bibinfo{volume}{348}
  (\bibinfo{year}{1994}) \bibinfo{pages}{177--181}.
\bibitem[{Zhao et~al.(1997)Zhao, Buchinger, Crawford, Fedrigo, Gulick, Lee,
  Constantinescu, Hussonnois, and Pinard}]{Zhao1997}
\bibinfo{author}{W.~Z. Zhao}, \bibinfo{author}{F.~Buchinger},
  \bibinfo{author}{J.~E. Crawford}, \bibinfo{author}{S.~Fedrigo},
  \bibinfo{author}{S.~Gulick}, \bibinfo{author}{J.~K.~P. Lee},
  \bibinfo{author}{O.~Constantinescu}, \bibinfo{author}{M.~Hussonnois},
  \bibinfo{author}{J.~Pinard}, \bibinfo{journal}{Hyperfine Interactions}
  \bibinfo{volume}{108} (\bibinfo{year}{1997}) \bibinfo{pages}{483--495}.
\bibitem[{Brown and Merer(1979)}]{Brown1979}
\bibinfo{author}{J.~M. Brown}, \bibinfo{author}{A.~J. Merer},
  \bibinfo{journal}{Journal of Molecular Spectroscopy} \bibinfo{volume}{74}
  (\bibinfo{year}{1979}) \bibinfo{pages}{488--494}.
\bibitem[{Brown et~al.(1987)Brown, Cheung, and Merer}]{Brown1987}
\bibinfo{author}{J.~Brown}, \bibinfo{author}{A.~S.-C. Cheung},
  \bibinfo{author}{A.~J. Merer}, \bibinfo{journal}{Journal of Molecular
  Spectroscopy} \bibinfo{volume}{124} (\bibinfo{year}{1987})
  \bibinfo{pages}{464--475}.
\bibitem[{Wang and Steimle(2012)}]{Wang2012}
\bibinfo{author}{F.~Wang}, \bibinfo{author}{T.~C. Steimle},
  \bibinfo{journal}{The Journal of Chemical Physics} \bibinfo{volume}{136}
  (\bibinfo{year}{2012}) \bibinfo{pages}{044312}.
\bibitem[{Klinkenberg et~al.(1961)Klinkenberg, {Van Kleef}, and
  Noorman}]{Klinkenberg1961}
\bibinfo{author}{P.~F.~A. Klinkenberg}, \bibinfo{author}{T.~A.~M. {Van Kleef}},
  \bibinfo{author}{P.~E. Noorman}, \bibinfo{journal}{Physica}
  \bibinfo{volume}{27} (\bibinfo{year}{1961}) \bibinfo{pages}{1177--1188}.
\bibitem[{Mulliken and Christy(1931)}]{Mulliken1931}
\bibinfo{author}{R.~S. Mulliken}, \bibinfo{author}{A.~Christy},
  \bibinfo{journal}{Physical Review} \bibinfo{volume}{38}
  (\bibinfo{year}{1931}) \bibinfo{pages}{87--119}.
\bibitem[{Mosyagin et~al.(2010)Mosyagin, Zaitsevskii, and Titov}]{Mosyagin:10a}
\bibinfo{author}{N.~S. Mosyagin}, \bibinfo{author}{A.~V. Zaitsevskii},
  \bibinfo{author}{A.~V. Titov}, \bibinfo{journal}{Review of Atomic and
  Molecular Physics} \bibinfo{volume}{1} (\bibinfo{year}{2010})
  \bibinfo{pages}{63--72}.
\bibitem[{Alekseyev et~al.(2004)Alekseyev, Liebermann, and
  Buenker}]{Alekseyev:04a}
\bibinfo{author}{A.~B. Alekseyev}, \bibinfo{author}{H.-P. Liebermann},
  \bibinfo{author}{R.~J. Buenker}, in: \bibinfo{editor}{K.~Hirao},
  \bibinfo{editor}{Y.~Ishikawa} (Eds.), \bibinfo{booktitle}{Recent Advances in
  Relativistic Molecular Theory}, \bibinfo{publisher}{World Scientific},
  \bibinfo{address}{Singapore}, \bibinfo{year}{2004}, pp.
  \bibinfo{pages}{65--105}.
\bibitem[{Titov et~al.(2001)Titov, Mosyagin, Alekseyev, and Buenker}]{Titov:01}
\bibinfo{author}{A.~V. Titov}, \bibinfo{author}{N.~S. Mosyagin},
  \bibinfo{author}{A.~B. Alekseyev}, \bibinfo{author}{R.~J. Buenker},
  \bibinfo{journal}{International Journal of Quantum Chemistry}
  \bibinfo{volume}{81} (\bibinfo{year}{2001}) \bibinfo{pages}{409--421}.
\bibitem[{Stanton et~al.(2011)Stanton, Gauss, Harding, Szalay et~al.}]{CFOUR}
\bibinfo{author}{J.~F. Stanton}, \bibinfo{author}{J.~Gauss},
  \bibinfo{author}{M.~E. Harding}, \bibinfo{author}{P.~G. Szalay}, et~al.,
  \bibinfo{title}{{``{\sc cfour}''}}, \bibinfo{year}{2011}. \bibinfo{note}{{\sc
  cfour}: a program package for performing high-level quantum chemical
  calculations on atoms and molecules, {http://www.cfour.de} .}
\bibitem[{Shavitt(1977)}]{Shavitt:77}
\bibinfo{author}{I.~Shavitt}, in: \bibinfo{editor}{H.~F. {Schaefer~III}} (Ed.),
  \bibinfo{booktitle}{Methods of Electronic Structure Theory},
  volume~\bibinfo{volume}{3} of \textit{\bibinfo{series}{Modern Theoretical
  Chemistry}}, \bibinfo{publisher}{Plenum Press}, \bibinfo{address}{New York},
  \bibinfo{year}{1977}, pp. \bibinfo{pages}{189--275}. \bibinfo{note}{462 p.}
\bibitem[{Roos et~al.(2005)Roos, Lindh, \r{A} Malmqvist, Veryazov, and
  Widmark}]{Roos:05}
\bibinfo{author}{B.~O. Roos}, \bibinfo{author}{R.~Lindh},
  \bibinfo{author}{P.~\r{A} Malmqvist}, \bibinfo{author}{V.~Veryazov},
  \bibinfo{author}{P.-O. Widmark}, \bibinfo{journal}{Journal of Physical
  Chemistry A} \bibinfo{volume}{108} (\bibinfo{year}{2005})
  \bibinfo{pages}{2851}.
\bibitem[{Petrov(2011)}]{Petrov:11}
\bibinfo{author}{A.~N. Petrov}, \bibinfo{journal}{Physical Review A}
  \bibinfo{volume}{83} (\bibinfo{year}{2011}) \bibinfo{pages}{024502}.
\bibitem[{Dudley and Coen(2006)}]{Dudley2006}
\bibinfo{author}{J.~M. Dudley}, \bibinfo{author}{S.~Coen},
  \bibinfo{journal}{Reviews of Modern Physics} \bibinfo{volume}{78}
  (\bibinfo{year}{2006}) \bibinfo{pages}{1135--1184}.
\bibitem[{Paul et~al.(2008)Paul, Johnson, Lee, and Jones}]{Paul2008}
\bibinfo{author}{J.~Paul}, \bibinfo{author}{J.~Johnson},
  \bibinfo{author}{J.~Lee}, \bibinfo{author}{R.~J. Jones},
  \bibinfo{journal}{Optics Letters} \bibinfo{volume}{33} (\bibinfo{year}{2008})
  \bibinfo{pages}{2482--4}.
\bibitem[{Tauser et~al.(2004)Tauser, Adler, and Leitenstorfer}]{Tauser2004}
\bibinfo{author}{F.~Tauser}, \bibinfo{author}{F.~Adler},
  \bibinfo{author}{A.~Leitenstorfer}, \bibinfo{journal}{Optics Letters}
  \bibinfo{volume}{29} (\bibinfo{year}{2004}) \bibinfo{pages}{516--8}.
\bibitem[{Cossel et~al.(2010)Cossel, Adler, Bertness, Thorpe, Feng, Raynor, and
  Ye}]{Cossel2010}
\bibinfo{author}{K.~C. Cossel}, \bibinfo{author}{F.~Adler},
  \bibinfo{author}{K.~A. Bertness}, \bibinfo{author}{M.~J. Thorpe},
  \bibinfo{author}{J.~Feng}, \bibinfo{author}{M.~W. Raynor},
  \bibinfo{author}{J.~Ye}, \bibinfo{journal}{Applied Physics B}
  \bibinfo{volume}{100} (\bibinfo{year}{2010}) \bibinfo{pages}{917--924}.
\bibitem[{Sun et~al.(2007)Sun, Gale, and Reid}]{Sun2007}
\bibinfo{author}{J.~H. Sun}, \bibinfo{author}{B.~J.~S. Gale},
  \bibinfo{author}{D.~T. Reid}, \bibinfo{journal}{Optics Letters}
  \bibinfo{volume}{32} (\bibinfo{year}{2007}) \bibinfo{pages}{1414--1416}.
\bibitem[{Gambetta et~al.(2008)Gambetta, Ramponi, and Marangoni}]{Gambetta2008}
\bibinfo{author}{A.~Gambetta}, \bibinfo{author}{R.~Ramponi},
  \bibinfo{author}{M.~Marangoni}, \bibinfo{journal}{Optics Letters}
  \bibinfo{volume}{33} (\bibinfo{year}{2008}) \bibinfo{pages}{2671--3}.
\bibitem[{Adler et~al.(2009)Adler, Cossel, Thorpe, Hartl, Fermann, and
  Ye}]{Adler2009}
\bibinfo{author}{F.~Adler}, \bibinfo{author}{K.~C. Cossel},
  \bibinfo{author}{M.~J. Thorpe}, \bibinfo{author}{I.~Hartl},
  \bibinfo{author}{M.~E. Fermann}, \bibinfo{author}{J.~Ye},
  \bibinfo{journal}{Optics Letters} \bibinfo{volume}{34} (\bibinfo{year}{2009})
  \bibinfo{pages}{1330--1332}.
\bibitem[{Leindecker et~al.(2011)Leindecker, Marandi, Byer, and
  Vodopyanov}]{Leindecker2011}
\bibinfo{author}{N.~Leindecker}, \bibinfo{author}{A.~Marandi},
  \bibinfo{author}{R.~L. Byer}, \bibinfo{author}{K.~L. Vodopyanov},
  \bibinfo{journal}{Optics Express} \bibinfo{volume}{19} (\bibinfo{year}{2011})
  \bibinfo{pages}{6296--302}.
\bibitem[{Neely et~al.(2011)Neely, Johnson, and Diddams}]{Neely2011}
\bibinfo{author}{T.~W. Neely}, \bibinfo{author}{T.~A. Johnson},
  \bibinfo{author}{S.~A. Diddams}, \bibinfo{journal}{Optics Letters}
  \bibinfo{volume}{36} (\bibinfo{year}{2011}) \bibinfo{pages}{4020--2}.
\bibitem[{Johnson and Diddams(2012)}]{Johnson2011a}
\bibinfo{author}{T.~A. Johnson}, \bibinfo{author}{S.~A. Diddams},
  \bibinfo{journal}{Applied Physics B} \bibinfo{volume}{107}
  (\bibinfo{year}{2012}) \bibinfo{pages}{31--39}.

\end{thebibliography}






\onecolumn
\ctable[
caption = 
{\textbf{Fitted constants for observed transitions in $^{180}$HfF$^+$ in cm$^{-1}$.} Quoted errors are at 95\% confidence level, statistical only. For bands recorded with the wavemeter uncalibrated we include an additional 0.006~\cm uncertainty in $\nu_0$. Internal consistencies in the data suggest that for the $\Delta v = 0$ bands, which typically suffer from partially resolved isotope structure, systematic errors in the fit quantities may exceed the quoted statistical uncertainties in some cases by a factor of two or three. Values without uncertainties were fixed in the fits. All constants are given for the vibrational levels involved in the transition and are not equilibrium values. The $\Lambda$-doubling terms, $k''$, $k'$, and $k_D'$ are generic terms proportional to $J(J+1)$ in the ground and excited states and $J^2(J+1)^2$ in the excited state respectively (see text for more details).} ,
label = transitions, pos = p]{lcccccccc} {

\tnote[${\circ}$] {$^{3}\Pi_{1}\leftarrow\!^{1}\Sigma^{+} (3,1)$ values assigned using a manual fit to multiple isotopes due to challenges of determining the line centers of the weaker lines in the dense spectrum.}
\tnote[*] {Value fixed to the fitted value from the $^1\Pi_1\!\leftarrow\!^1\Sigma^+$ (0, 1) transition.}
\tnote[$+$] {Individual isotopes were not fully resolved, thus the fit was done to averaged line positions. We report isotope corrected values with the error due to the uncertainty in the isotope shifts.}
\tnote[$\dagger$] {$^3\Phi_2\!\leftarrow\!^3\Delta_1$ transitions exhibited line-doubling at high J'', but the doublets were not resolved well enough to accurately determine k''.} 
\tnote[$^\ddagger$] {Each transition was a doublet with a splitting given by $k' \times J'(J'+1)$ due to $\Lambda$-doubling in $^1\Pi_1$.}
}{
	\FL
									&  $\nu_0$ 			&  $B'' $		& $B' $		&  $D'' $		&  $D' $		& $k'' $		& $k' $		& $k_D'$ \NN
									& 					& 			& 			& [$10^{-7}$]	& [$10^{-7}$]	& [$10^{-4}$]	& [$10^{-4}$]	& [$10^{-9}$] \ML
$^1\Pi_1\!\leftarrow\!^1\Sigma^+$ (0, 1)		& 12217.369(2)		& 0.30335(2)	& 0.28115(3)	& 1.88(8)		& 1.81(8)		& \textendash\	& 3.69(2)		& 9.7(7) \NN
$^1\Pi_1\!\leftarrow\!^1\Sigma^+$ (1, 2)		& 12136.012(3)		& 0.30180(5)	& 0.27973(5)	& 1.80(14)	& 1.74(14)	& \textendash\	& 2.68(4)		& 7.4(1.3) \NN
$^1\Pi_1\!\leftarrow\!^1\Sigma^+$ (0, 0)		& 13002.189(12)$^+$	& 0.30474(20)	& 0.28104(20)	& 1(2)		& 1(2)		& \textendash\	& 3.55(4)		& 0 \NN
$^3\Pi_1\!\leftarrow\!^1\Sigma^+$ (2, 0)		& 12304.400(3)		& 0.30481(5)	& 0.28096(5)	& 1.78(12)	& 1.78(12)	& \textendash\	& -3.82(1)		& 0 \NN
$^3\Pi_1\!\leftarrow\!^1\Sigma^+$ (3, 1)$^{\circ}$	& 12216.901(10)	& 0.30335*	& 0.27958(20)	& 1.8(3)		& 1.8(3)		& \textendash\	& -3.8(2)		& 0 \NN
$^3\Pi_1\!\leftarrow\!^1\Sigma^+$ (0, 1)		& 10109.877(6)		& 0.30333(6)	& 0.28382(6)	& 1.9(5)		& 1.9(5)		& \textendash\	& -3.77(5)		& 0 \NN
$^3\Pi_{0^+}\!\leftarrow\!^1\Sigma^+$ (0, 0)	& 10401.723(13)$^+$	& 0.30500(10)	& 0.28437(11)	& 2.3(1.2)		& 2.2(1.2)		& \textendash\	& \textendash\	&  \textendash \NN
$^3\Pi_{0^+}\!\leftarrow\!^1\Sigma^+$ (1, 0)	& 11114.653(7)		& 0.30482(4)	& 0.28281(4)	& 1.71(14)	& 1.72(13)	& \textendash\	& \textendash\	&  \textendash \NN
$^3\Sigma_{0^+}^-\!\leftarrow\!^1\Sigma^+$ (0, 0)	& 13254.302(7)	& 0.30478(10)	& 0.28967(9)	& 1.7(5)		& 1.9(5)		& \textendash\	& \textendash\	&  \textendash \NN
$^3\Sigma_{0^+}^-\!\leftarrow\!^1\Sigma^+$ (1, 0)	& 13953.799(6)	& 0.30483(5)	& 0.28808(4)	& 1.78(12)	& 1.98(13)	& \textendash\	& \textendash\	&  \textendash \NN
$^3\Phi_2\!\leftarrow\!^3\Delta_1$ (0, 0)		& 13933.340(12)$^+$	& 0.29900(10)	& 0.27720(10)	& 2.2(3)		& 2.1(3)		& $\dagger$	& -0.01(3)		&  \textendash \NN
$^3\Phi_2\!\leftarrow\!^3\Delta_1$ (1, 0)		& 14613.782(7)		& 0.29888(5)	& 0.27590(5)	& 1.71(15)	& 1.59(15)	& $\dagger$	& 0			&  \textendash \NN
$^3\Phi_2\!\leftarrow\!^3\Delta_1$ (2, 1)		& 14534.839(7)		& 0.29745(8)	& 0.27488(8)	& 1.7(4)		& 1.6(4)		& $\dagger$	& 0			&  \textendash \NN
$^3\Pi_{0^+}\!\leftarrow\!^3\Delta_1$ (1, 0)	& 10137.723(7)		& 0.29887(5)	& 0.28277(5)	& 1.70(20)	& 1.66(20)	& -0.14(2)		& \textendash\	&  \textendash \NN
$^3\Pi_{0^-}\!\leftarrow\!^3\Delta_1$ (1, 0)	& 9948.624(6)			& 0.29891(2)	& 0.28272(2)	& 1.85(5)		& 1.81(4)		& -0.123(6)	& \textendash\	&  \textendash \NN
$^1\Pi_1\!\leftarrow\!^3\Delta_2$ (0, 0)		& 10852.757(13)$^+$	& 0.29915(9)	& 0.28102(9)	& 1.7(4)		& 1.7(4)		& 0			& 3.59(3)$^\ddagger$&  \textendash \NN
?	$\Delta v = 0$						&13729.918(17)$^+$	& 0.29860(60)	& 0.28680(60)	& 3(3)		& 3(3)		& 0			& -0.2(3)		&  \textendash \NN
?	$\Delta v = 0$						& 13822.270(14)$^+$	& 0.29980(10)	& 0.28860(10)	& 1.9(2)		& 2.2(2)		& 0			& 0			&  \textendash \NN
?	$\Delta v = 0$						& 14636.162(13)$^+$	& 0.29620(30)	& 0.27600(30)	& 1.7(1.5)		& 1.8(1.5)		& 0			& 0			&  \textendash \LL
}


\ctable[
caption = {\textbf{Derived constants for observed states in  $^{180}$HfF$^+$ in cm$^{-1}$.} Values for T$_0$ are given to the $\nu=0$ levels, with the X$^1\Sigma^+$ $\nu=0$ level set to 0. Values for T$_e$ are to the minimum of the potential curves again with X$^1\Sigma^+$ set to 0. $\Delta$G$_{1/2}$ is only given for states where we had a direct measurement. All equilibrium values assume a Morse potential unless otherwise noted. Quoted uncertainties are 95\% and are statistical, i.e., they do not account for any deviation from the Morse potential. $\delta T_e = T_e^{180} - T_e^{178}$ is an electronic isotope shift relative to $^1\Sigma^+$ for each state as discussed in the text.},
label = stateconstants, pos = hb
]{lcccccccc}{
\tnote[${+}$] {Calculated without assuming a Morse potential.} 
\tnote[*] {Assuming $\omega_e$, $\omega_e x_e$, and $\alpha_e$ are the same as for the $^{3}\Pi_{0+}$.}
}{
	\FL
				& T$_0$ 			& $\Delta$G$_{1/2}$		& T$_e$			&  B$_e$ 		&  $\omega_e$ 	& $\omega_e x_e$ 	& $\alpha_e$ [$10^{-3}$]	& $\delta$T$_e$\ML
$^1\Sigma^+$		& 0				& 784.820(12)			& 0				& 0.30558(3)		& 790.76(11)		&  2.97(5)			& 1.50(2)			& 0\NN
$^{3}\Delta_{1}$	& 976.930(10)		& \textendash\			& 991.83(74)		& 0.29963(5)		& 760.9(1.5)		&  2.78(21	)		& 1.45(8)			& -0.049(12) \NN
$^{3}\Delta_{2}$	& 2149.432(18)	& \textendash\			& \textendash\		& \textendash\		& \textendash\		&  \textendash\		& \textendash\		& -0.055(12) \NN
$^1\Pi_1$			& 13002.189(12)	& \textendash\			& 13046.04(21)	& 0.28186(5)		& 702.9(4)		&  2.70(15)		& 1.42(6)			& -0.071(6) \NN
$^3\Pi_1$			& 10894.697(14)	& \textendash\			& 10933.77(6)$^+$	& 0.28454(8)		& 712.382(21)$^+$	&  2.51(6)$^+$		& 1.43(4)			& -0.054(7) \NN
$^{3}\Pi_{0^-}$		& \textendash		& \textendash\			& 10248.34(36)*	& 0.28517(17)*		& \textendash\		&  \textendash\		& \textendash\		& -0.033(16) \NN
$^{3}\Pi_{0^+}$		& 10401.723(13)	& 712.930(15)			& 10437.44(36)	& 0.28518(13)		& 719.5(7)		&  3.3(3)			& 1.63(11)		& -0.047(7) \NN
$^{3}\Sigma_{0^+}^-$	& 13254.302(7)	& 699.497(9)			& 13296.93(28)	& 0.29046(14)		& 705.5(5)		&  3.0(3) 			& 1.59(10)		& -0.106(7) \NN
$^{3}\Phi_{2}$		& 14910.270(16)	& 680.442(14)			& 14963.15(33)	& 0.27778(14)		& 684.5(7)		&  2.05(31)		& 1.16(14)		& -0.044(16)\LL
	}	

\ctable[
caption = {\textbf{Summary of states.} Comparison of derived molecular constants from this work and from the experiments of Barker et al. \cite{Barker2011} with the old theoretical calculations of Petrov et al. \cite{Petrov2009} and the improved theory discussed here. The theoretical values of $B_{e}$ from \cite{Petrov2009} were computed from the equilibrium bond length. Experimental uncertainties are converted to 95\% (2$\sigma$).},
label = theorycomp, pos = p
]{lccccc} {
\tnote[$+$] {$\Delta G_{1/2}$.}
\tnote[$\dagger$] {T$_0$.}
\tnote[*] {$B_0$.}
}{
\FL
	State		&	Constant			&	This Work				&	 \cite{Barker2011}		&  \cite{Petrov2009} 		&   New theory \ML
$^{1}\Sigma^{+}$ &	T$_e$			&	0					&	0					&	0				&	0 	\NN
			&	B$_e$			&	0.30558(3)			&	0.304(10)				&	0.3082			&	0.309	\NN
			&	$\omega_e$		&	790.76(11)			&	791.2(1.0)				&	751				&	792 	\NN
\NN
$^{3}\Delta_{1}$ 	&	T$_e$		&	991.83(74)	 		&	993(2)				&	1599				&	1229 	\NN
			&	B$_e$			&	0.29963(5)			&	0.301(10)				&	0.2994			&	0.301	\NN
			&	$\omega_e$		&	760.9(1.5)				&	761.3(2.0)				&	718				&	754 	\NN
\NN
$^{3}\Delta_{2}$ 	&	T$_e$		&	2149.432(16)$^\dagger$	&	2151.7(20)$^\dagger$	&	2807				&	2394 	\NN
			&	B$_e$			&	0.29915(9)*			&	0.300(10)				&	0.2997			&	0.302	\NN
			&	$\omega_e$		&	\textendash			&	762.3(2.0)				&	719				&	766 	\NN
\NN
$^{3}\Delta_{3}$ 	&	T$_e$		&	\textendash 			&	3951(2)				&	4324				&	3995 	\NN
			&	B$_e$			&	\textendash			&	0.308(10)				&	0.3004			&	0.301	\NN
			&	$\omega_e$		&	\textendash			&	761.5(2.0)$^+$			&	721				&	757 	\NN
\NN
$^{1}\Delta_{2}$ 	&	T$_e$		&	\textendash			&	\textendash 			&	11519			&	10610 	\NN
			&	B$_e$			&	\textendash			&	\textendash			&	0.2981			&	0.298	\NN
			&	$\omega_e$		&	\textendash			&	\textendash 			&	696				&	747 	\NN
\NN
$^{3}\Pi_{0^-}$ 	&	T$_e$			&	10248.34(36)			&	\textendash 			&	11910			&	10400 	\NN
			&	B$_e$			&	0.28517(17)			&	\textendash 			&	0.2848			&	0.286	\NN
			&	$\omega_e$		&	\textendash			&	\textendash 			&	689				&	716 	\NN
\NN
$^{3}\Pi_{0^+}$	&	T$_e$			&	10437.44(36)			&	\textendash 			&	12196			&	10658 	\NN
			&	B$_e$			&	0.28518(13)			&	\textendash 			&	0.2854			&	0.285	\NN
			&	$\omega_e$		&	719.5(7)				&	\textendash 			&	699				&	724 	\NN
\NN
$^{3}\Pi_{1}$ 	&	T$_e$			&	10933.77(6)			&	\textendash 			&	12686			&	11058 	\NN
			&	B$_e$			&	0.28454(8)			&	\textendash 			&	0.2835			&	0.285	\NN
			&	$\omega_e$		&	712.382(17)			&	\textendash 			&	687				&	712 	\NN
\NN
$^{3}\Pi_{2}$ 	&	T$_e$			&	\textendash			&	\textendash 			&	14438			&	13452 	\NN
			&	B$_e$			&	\textendash			&	\textendash 			&	0.2848			&	0.287	\NN
			&	$\omega_e$		&	\textendash			&	\textendash 			&	703				&	745 	\NN
\NN
$^{1}\Pi_{1}$ 	&	T$_e$			&	13046.3(3)			&	\textendash 			&	14784			&	13493 	\NN
			&	B$_e$			&	0.28186(5)			&	\textendash 			&	0.2805			&	0.283	\NN
			&	$\omega_e$		&	702.9(4)				&	\textendash 			&	679				&	699 	\NN
\NN
$^{3}\Sigma_{0^+}^-$ 	&	T$_e$	&	13296.93(28)			&	\textendash\ 			&	\textendash\ 		&	13773 	\NN
			&	B$_e$			&	0.29046(14)			&	\textendash\ 			&	\textendash\ 		&	0.292	\NN
			&	$\omega_e$		&	705.5(5)				&	\textendash\ 			&	\textendash\ 		&	716 	\LL			
}

\ctable[
caption = {\textbf{Summary of states.}},
 pos = p, continued
]{lccccc} {
\tnote[${\circ}$] {Estimated from $\Lambda$-doubling in the $^{3}\Pi_{0}$ (see text).}
}{
\FL
	State		&	Constant			&	This Work				&	 \cite{Barker2011}		&  \cite{Petrov2009} 		&   New theory \ML
$^{3}\Sigma_{1}^-$ 	&	T$_e$		&	\textendash\ 			&	\textendash\ 			&	\textendash\ 		&	14757 	\NN
			&	B$_e$			&	\textendash\			&	\textendash\ 			&	\textendash\ 		&	0.292	\NN
			&	$\omega_e$		&	\textendash\			&	\textendash\ 			&	\textendash\ 		&	711 	\NN
\NN
$^{3}\Phi_{2}$ 	&	T$_e$			&	14963.15(33)			&	\textendash\ 			&	\textendash\ 		&	15284	\NN
			&	B$_e$			&	0.27778(14)			&	\textendash\ 			&	\textendash\ 		&	0.278	\NN
			&	$\omega_e$		&	684.5(7)				&	\textendash\ 			&	\textendash\ 		&	671 	\NN
\NN
$^{3}\Phi_{3}$ 	&	T$_e$			&	\textendash			&	\textendash\ 			&	\textendash\ 		&	17457	\NN
			&	B$_e$			&	\textendash			&	\textendash\ 			&	\textendash\ 		&	0.277	\NN
			&	$\omega_e$		&	\textendash			&	\textendash\ 			&	\textendash\ 		&	658 	\NN
\NN
$^{1}\Gamma_{4}$ 	&	T$_e$		&	\textendash			&	\textendash\ 			&	\textendash		&	18312 	\NN
			&	B$_e$			&	\textendash			&	\textendash\ 			&	\textendash		&	0.289	\NN
			&	$\omega_e$		&	\textendash			&	\textendash\ 			&	\textendash		&	641 	\NN
\NN
$^{3}\Pi_{0^-}$ 	&	T$_e$			&	\textendash			&	\textendash\ 			&	\textendash		&	19167 	\NN
			&	B$_e$			&	\textendash			&	\textendash\ 			&	\textendash		&	0.276	\NN
			&	$\omega_e$		&	\textendash			&	\textendash\ 			&	\textendash		&	691 	\NN
\NN
$^{3}\Pi_{1}$ 	&	T$_e$			&	\textendash			&	\textendash\ 			&	\textendash		&	19332 	\NN
			&	B$_e$			&	\textendash			&	\textendash\ 			&	\textendash		&	0.279	\NN
			&	$\omega_e$		&	\textendash			&	\textendash\ 			&	\textendash		&	698 	\NN
\NN
$^{3}\Pi_{0^+}$	&	T$_e$			&	\textendash			&	\textendash\ 			&	\textendash		&	20074 	\NN
			&	B$_e$			&	\textendash			&	\textendash\ 			&	\textendash		&	0.280	\NN
			&	$\omega_e$		&	\textendash			&	\textendash\ 			&	\textendash		&	748 	\NN
\NN
$^{1}\Sigma^{+}$ 	&	T$_e$		&	\textendash			&	\textendash\ 			&	\textendash\ 		&	20330	\NN
			&	B$_e$			&	\textendash			&	\textendash\ 			&	\textendash\ 		&	0.288	\NN
			&	$\omega_e$		&	\textendash			&	\textendash\ 			&	\textendash\ 		&	610 	\NN
\NN
$^{3}\Pi_{2}$ 	&	T$_e$			&	\textendash			&	\textendash\ 			&	\textendash		&	20338 	\NN
			&	B$_e$			&	\textendash			&	\textendash\ 			&	\textendash		&	0.277	\NN
			&	$\omega_e$		&	\textendash			&	\textendash\ 			&	\textendash		&	665 	\NN
\NN
$^{3}\Phi_{4}$ 	&	T$_e$			&	\textendash			&	\textendash\ 			&	\textendash\ 		&	20769	\NN
			&	B$_e$			&	\textendash			&	\textendash\ 			&	\textendash\ 		&	0.283	\NN
			&	$\omega_e$		&	\textendash			&	\textendash\ 			&	\textendash\ 		&	740 	\NN
\NN
$^{3}\Sigma_{1}^+$ 	&	T$_e$		&	\textendash			&	\textendash\ 			&	\textendash\ 		&	21415	\NN
			&	B$_e$			&	\textendash			&	\textendash\ 			&	\textendash\ 		&	0.276	\NN
			&	$\omega_e$		&	\textendash			&	\textendash\ 			&	\textendash\ 		&	665 	\NN
\NN
$^{3}\Sigma_{0^-}^+$ 	&	T$_e$	&	21200$^{\circ}$		&	\textendash\ 			&	\textendash\ 		&	21694	\NN
			&	B$_e$			&	\textendash			&	\textendash\ 			&	\textendash\ 		&	0.277	\NN
			&	$\omega_e$		&	\textendash			&	\textendash\ 			&	\textendash\ 		&	658 	\LL
}


\end{document}